\documentclass{aa}  
\usepackage{amsmath}
\usepackage[colorinlistoftodos]{todonotes}
\usepackage{natbib}
\bibpunct{(}{)}{;}{a}{}{,} 
\usepackage[T1]{fontenc}
\usepackage[english]{babel} 
\usepackage{amssymb}
\usepackage{mathrsfs}
\usepackage{soul} 
\usepackage{ulem} 
\usepackage{multicol,float} 
\usepackage{caption}
\usepackage{textcomp} 
\usepackage[pdftex,colorlinks=true,linkcolor=black,citecolor=blue,urlcolor=blue]{hyperref}
\everymath{\displaystyle} 
\usepackage{tabularx}
\newcolumntype{C}{>{\centering}X}
\usepackage[version=4]{mhchem}

\usepackage{subcaption}
\usepackage{gensymb}
\usepackage{graphicx}

\usepackage{txfonts}

\usepackage{blindtext}

\makeatletter
\renewcommand{\paragraph}{\@startsection{paragraph}{4}{0ex}%
	{-3.25ex plus -1ex minus -0.2ex}%
	{1.5ex plus 0.2ex}%
	{\normalfont\normalsize}}
\makeatother

\stepcounter{secnumdepth}
\stepcounter{tocdepth}
\begin{document}

	\title{OH mid-infrared emission as a diagnostic of H$_2$O UV photodissociation.}
	
	\subtitle{II. Application to interstellar PDRs}
	
	\author{M. Zannese
		\inst{1}, B. Tabone\inst{1},
		E. Habart\inst{1}, F. Le Petit\inst{2}, E. van Dishoeck\inst{3,4}, E. Bron\inst{2}}
	\authorrunning{M. Zannese
		\inst{1}}
	
	\institute{Université Paris-Saclay, CNRS, Institut d'Astrophysique Spatiale, 91405 Orsay\\
		\email{marion.zannese@universite-paris-saclay.fr} \and
		LERMA,Observatoire de Paris, PSL Research University, CNRS, Sorbonne Universités, 92190 Meudon, France \and Leiden Observatory, Leiden University, 2300 RA Leiden, The Netherlands \and Max-Planck Institut für Extraterrestrische Physik (MPE), Giessenbachstr. 1, 85748 Garching, Germany}
	
	\date{25th of August 2022}
	
	
	\abstract
	{Water photodissociation in the $114-143~$nm UV range forms excited OH which emits at mid-infrared wavelengths via highly excited rotational lines. These lines have only been detected with \textit{Spitzer} in several proto-planetary disks and shocks. Previous studies have shown they are a unique diagnostic for water photodissociation. Thanks to its high sensitivity and angular resolution, the \textit{James Webb Space Telescope (JWST)} could be able to detect them in other environments such as interstellar Photo-Dissociation Regions (PDRs).}
	{Our goal is to predict OH mid-IR lines for a large range of thermal pressures and UV fields in PDRs.}
	{We use the Meudon PDR Code to compute the thermal and chemical structure of PDRs.
		In order to predict the emerging spectrum of OH, we amended the code to include prompt emission induced by H$_2$O photodissociation between $114$ and $143$ nm. The influence of thermal pressure ($P_{\rm th}/k$ = $n_{\rm H} T_{\rm K}$) and UV field strength on the integrated intensities, as well as their detectability with the \textit{JWST} are studied in details.}
	{OH mid-IR emission is predicted to originate very close to the H$^0$/H$_2$ transition and is directly proportional to the column density of water photodissociated in that layer. Because gas-phase neutral-neutral reactions forming water require relatively high temperatures ($T_{\rm K} \gtrsim 300~$K), the resulting OH mid-IR lines are primarily correlated with the temperature at this position, and are therefore brighter in regions with high pressure. This implies that these lines are predicted to be only detectable in strongly irradiated PDRs ($G_0^{\rm incident}$ $>$ 10$^3$) with high thermal pressure ($P_{\rm th}/k$ $\gtrsim$ 5$\times$10$^7$ K cm$^{-3}$). In the latter case, OH mid-IR lines are less dependent on the strength of the incident UV field. The detection in PDRs like the Orion bar, which should be possible, is also investigated and we show that the line-to-continuum ratio can be a major limitation for detection because of instrumental limitations.}
	{OH mid-IR lines observable by \textit{JWST} are a promising diagnostics for dense and strongly irradiated PDRs and proplyds. Their intensities are directly proportional the amount of water photodissociated and therefore an indirect but sensitive probe of the gas temperature at the H$^0$/H$_2$ transition.} 
	
	
	
	\maketitle
	
	\section{Introduction}
	
	Photo-Dissociation Regions (PDRs) are the place where the radiative feedback is dominant, with intense stellar Far-UV (FUV) radiation coming from stars in galaxies playing a dominant role in the physics and chemistry \citep[for a review, see for example,][]{Hollenbach_1999,Wolfire_2022}. Stellar feedback is one of the major mechanisms to limit star formation \citep[e.g.,][]{Inoguchi_2020} by contributing to dispersal of the cloud due to gas heating and angular momentum addition. The study of these regions is then essential to have a better understanding on star formation and evolution of the interstellar matter. As a consequence of the intense UV field \citep[up to a few 10$^5$ in units of the Mathis field corresponding to 1.9 $\times$ 10$^{-3}$ erg s$^{-1}$ cm$^{-2}$,][]{Mathis_1983}, PDRs act as a cradle of a very active chemistry which is even more enhanced in dense star forming regions (with density around $n_{\rm H}$ $\sim$ 10$^6$ cm$^{-3}$). Moreover, emission from PDRs which reprocess a significant part of the radiation energy emitted by young stars dominates the infrared spectra in the galaxy. It is therefore crucial to understand how the observed mid-IR emission is linked with physical conditions, and how these observations can constrain astrophysical environments.
	
	The study of interstellar PDRs such as the Orion Bar, NGC 7023 or the Horsehead nebula is also key for the understanding of the still unknown UV driven processes in other inter- and circumstellar media. Planetary nebulae observations and models have indicated that a large fraction of the gas ejected by evolved stars goes through a PDR phase before being injected in the interstellar medium (ISM) \citep{Hollenbach_1995,Bernard-Salas_2005}.  IR spectroscopy can give information on the initial physical and chemical properties of the PDR phase \citep[e.g.,][]{Bernard-Salas_2009,Cox_2016} and probe the photo-chemical evolution of molecules, nano-particles, and grains. Dense and highly irradiated PDRs are also present in the FUV-illuminated surfaces of protoplanetary disks \citep[e.g.,][]{Visser_2007,Woitke_2009}. 
	To conclude, PDRs are present in a wide variety of environments and the interstellar PDRs represent a unique laboratory to study UV driven micro-processes. 
	
	Because of the layered structure with a strong variation of physical conditions across the PDR, a multi-wavelength study is needed to trace every zone.  Emission lines with high upper level energy from atomic ions \citep[e.g.,  S$^+$, Si$^+$, Fe$^+$,][]{Kaufman_2006} trace the ionization front. Ro-vibrational and pure rotational lines of H$_2$ \citep[e.g.,][]{Parmar_1991,Luhman_1994,van_der_Werf_1996, Walmsley00, Allers_2005, Habart_2005,Habart_2011,Sheffer_2011,Kaplan_2017,Zhang_2021} probe the H$^0$/H$_2$ dissociation front. A bit further into the PDRs, the excited molecular gas is traced by rotationally excited lines from species such as CO \citep[e.g.,][]{Stutzki_1990,Tauber_1994,Hogerheijde_1995,Nagy_2017,Joblin_2018,Parikka_2018}, CH$^+$ or OH \citep[e.g.,][]{Goicoechea_2011, Parikka_2017}. Key PDR signatures in the infrared also include aromatic bands and
	dust continuum emission \citep[e.g.,][]{Tielens_1993,Compiegne_2011,Schirmer_2020}.

	In order to spatially resolve the scale of far-UV photon penetration (i.e., $A_V\sim$ 1) in dense molecular clouds, high angular resolution observations are needed. For instance, observations with the Atacama Large Millimeter/submillimeter Array (ALMA) with a spatial resolution of 1" have revealed a very complex  structure at the interface of the molecular cloud and the ionized gas for the highly illuminated Orion Bar PDR  instead of an homogenous layered structure \citep{Goicoechea_2016}. Moreover, recent NIR Keck/NIRC2 observations with a resolution of $\sim$0.1$''$ have resolved the sub-structures of this interface and especially at the H$^0$/H$_2$ transition with several ridges and extremely sharp filaments \citep{Habart_2022}. We observe a spatial coincidence between the H$_2$ 1-0 S(1) vibrational and HCO$^+$ J=4-3 rotational emission previously obtained with ALMA. This highlights that in high pressure PDR the H$^0$/H$_2$ and C$^+$/C$^0$/CO transition zones nearly coincide and are closer than expected for a typical layered structure of a constant density PDR. The need of angular resolution is even more important to resolve these two spatially close transitions. \\
	
	Up to now, the spatial resolution attained in the mid-IR domain (\textit{Spitzer}, \textit{ISO}, ...) have been too limited to resolve the sharp transition between hot ionized diffuse gas and cold molecular dense gas. The upcoming \textit{James Webb Space Telescope (JWST)} observations will improve our understanding of these regions as it will grant us new data in the IR wavelengths between 0.6 and 28.8 $\mu$m. In particular, the MIRI instrument of the \textit{JWST}, observing in the mid-IR, will combine high angular resolution maps (up to 0.2") and IFU spectroscopy giving access to spatially resolved spectra at each pixel of the map. Its high sensitivity will also enable the detection of numerous weak lines. Overall, the \textit{JWST} observations will permit to trace the warm and hot gas at small spatial scales and shed light on the FUV driven chemistry and the physical conditions in these regions. In particular, a well-observed highly illuminated PDR, the Orion Bar, will be the target of an Early Release Science PDRs4All (1288) observation with the \textit{JWST} \citep{ERS_2022} in addition to a GTO “Physics and Chemistry of PDR Fronts" (1192) focusing on the Horsehead nebula and NGC-7023. \\
	
	Among all the different lines in the mid-IR, the rotationaly excited OH lines in the $\lambda = 9-15$ $\mu$m range appear as a promising diagnostic to unveil UV driven processes in PDRs. Indeed, the incident UV field on the cloud leads to H$_2$O photodissociation that produces mostly OH with various quantum states. Interestingly, when photodissociation is caused by short wavelength photons (114 $\le$ $\lambda$ $\le$ 143 nm, i.e. via the $\tilde{B}$ state of H$_2$O), OH is formed in highly rotationally excitated states \citep[$\sim$ 40,000 K, corresponding to $N > 35$,][]{van_Harrevelt_2000}. The subsequent de-excitation of these nascent OH products via a radiative cascade produces mid-IR lines, a process called "prompt emission". Further modeling with a single-zone model, shows that the line fluxes give unique access to the amount of water photodissociated per unit time \citep{Tabone_2021}.  Therefore, in PDRs, mid-IR OH lines could be a unique diagnostic to constrain the physical conditions upon which the amount of photodissociated water is sensitive to, typically the temperature at the H$^0$/H$_2$ transition. Although previous studies have derived H$_2$O column densities using emission lines observed by \textit{Herschel} \citep[e.g. ][]{Choi_2014,Putaud_2019}, these data mostly probe the cold regions of the PDR where H$_2$O is very weakly photodissociated. Hence these results give access to different constraints than the OH mid-IR emission.
	Up to now, these rotationally excited lines of OH have however only been detected with \textit{Spitzer} in proto-planetary disks and strong protostellar shocks \citep[e.g.,][]{Tappe_2008,Tappe_2012,Carr_2014}. In PDRs and proplyds, only less excited far-IR lines of OH ($E_{\rm u}/k$ $\sim$ 100-300 K), more likely excited by collisions have been detected \citep[e.g.,][]{Goicoechea_2011,Parikka_2018}.  Therefore, the detectability and the potential of mid-IR lines of OH in PDRs and requires the use of detailed modeling.
	
	In this paper, we predict OH mid-IR line intensities in PDRs by computing in a consistent way the chemistry, thermal balance, and excitation of OH using the Meudon PDR code. We then study how the key physical parameters of a PDR, that are, in our modeling framework, the thermal pressure and the strength of the incident UV field, affect the intensities. The paper is organized as follows. In Sect. \ref{models}, we summarize the main ingredients of the Meudon PDR code and the updates made to model OH promt emission. In Sect. \ref{Results}, we present our main results on H$_2$O and OH chemistry and on the resulting mid-IR OH lines for a grid of models. In Sect. \ref{Orion_bar}, our model is applied to the Orion Bar where we discuss their detectability with \textit{JWST} taking into account continuum, bands and other lines. Finally, we discuss their detectability in other environments. Our findings are summarized in Sect. \ref{conclusion}.
	
	\section{Models}
	\label{models}
	\subsection{Thermo-chemical model with the Meudon PDR Code}
	
	\label{Meudon_PDR}
	
	In this work, we computed H$_2$O and OH density profiles and the local UV field intensity as a function of the depth into the PDR using the Meudon PDR code \citep[version 1.5.4\footnote{Public version of August 2021 : \url{https://pdr.obspm.fr/pdr\_download.html}},][]{Le_Petit_2006}. The code simulates in a self-consistent manner the thermal and chemical structure of the gas considering a 1D geometry and a stationary state in a plane-parallel irradiated gas and dust layer. This code takes as an input the shape of the incident UV field. Here, we use the \cite{Mathis_1983} prescription for the grid of models (see Sect. \ref{ref_model}), and a radiation field representative of an O7 star for our application to the Orion Bar (see Sect. \ref{Orion_bar}). The code includes the progressive attenuation of the UV field due to grain and gas extinction. In this work, we use a mean galactic extinction curve with the parameterization of \cite{Fitzpatrick_1988} for the grid of models (see Sect. \ref{ref_model}) and a flater extinction curve for the Orion Bar (see Sect. \ref{Orion_bar} for further details). The chemistry of the PDR is computed taking into account hundreds of species and thousands of chemical reactions. The excitation of the several key species is considered in the calculation of the thermal balance as the cooling relies on line emission. Photoelectric effect and thermal coupling between gas and dust are also taken into account in the thermal balance.
	
	In this paper, we assume that the PDR is isobaric. This is an appropriate starting hypothesis since several studies show that isobaric models reproduce the observed emission of warm molecular gas in interstellar PDRs such as the Horsehead, NGC-7023 and the Orion Bar \citep[e.g.,][]{Habart_2005,Allers_2005,Joblin_2018}. However,  magnetic and turbulence pressure may be important in PDRs \citep[e.g.,][]{vanDishoeck1986,Pellegrini09,Pabst20} and the thermal pressure might not dominate. For instance, in the Orion Bar, the non-thermal turbulent pressure is of the same order to the gas thermal pressure \citep[see Table 1. of ][]{Goicoechea_2016}. Consequently, the density and temperature gradients calculated in isobaric models may not fit fully the PDRs gas structure. However, in order to estimate the emission of mid-IR OH lines and investigate how it varies with the excitation and physical conditions (i.e., temperature, density), this assumption is valid as a first approximation. This study is based on a grid of models with thermal pressure ranging from $P_{\rm th}/k$ = 10$^5$ to 10$^9$ K cm$^{-3}$ ($P_{\rm th}/k$ = $n_{\rm H} T_{\rm K}$) and an intensity of the FUV field from $G_0^{\rm incident}$ = 10$^2$ to 10$^5$ in units of the interstellar radiation field of \cite{Mathis_1983}. 
	All the parameters necessary to the models are summarized in Table \ref{table:inputs}. A fiducial model with $P_{\rm th}/k$ = 10$^8$ K cm$^{-3}$ and $G_0^{\rm incident}$ = 10$^4$ (typical parameters of highly excited PDRs such as the Orion Bar) is adopted to present the results. In the following paragraphs, we review the micro-physical processes that are key for the modeling of the warm molecular layer where OH mid-IR emission originates.
	
	In the Meudon PDR Code, the H$_2$ rotational and rovibrational levels are calculated including collisional (de-)excitation with H \citep{Wrathmall_2007}, He, H$_2$ \citep{Flower_1998,Flower_1999} and H$^+$, and UV radiative pumping of electronic lines followed by fluorescence. For our grid of model, we use the FGK approximation \citep{Federman_1979} which allows a rapid computation for the UV radiative transfer involving self-shielding effects. The levels of H$_2$ are populated also considering excitation due to formation on grain surfaces. For the formation on dust surface, we assume a Boltzmann distribution at a temperature of 1/3 of H$_2$ dissociation energy \citep{Black_1987}. The
	distribution is uneven and probably depends on conditions in the PDR and the nature of the grains. As the branching ratio is unknown, it is assumed that distribution follows an equipartition law. The two other third of H$_2$ formation energy are distributed between grain excitation and kinetic energy of released molecules.

	The version of the code used in this paper includes an extensive chemical network. However, H$_2$O molecules formation is computed only taking into account the gas-phase chemistry and the formation on grains is neglected. As this study focuses on the photodissociation of water in warm region, the only formation mechanism interesting here is gas-phase formation so it does not affect the results. Chemical reaction rates are computed using thermal rate coefficients, except for the formation of CH$^+$, SH$^+$, OH and H$_2$O. For instance, a state-specific chemistry is included for the formation of OH and H$_2$O as a result of reactions with H$_2$ with energy barrier (see further in Sect. \ref{OH_H2O} with equations \ref{OH_form} and \ref{H2O_form}). This allows the internal energy of H$_2$ to be considered in the rate coefficient and the internal energy of H$_2$ may be used to overcome an activation barrier. Regarding the reaction OH+H$_2$(v,J), the state-specific chemistry is taken into account by replacing the activation energy by the difference between the activation energy and the ro-vibrational energy of H$_2$ \citep[e.g. ][]{Tielens_1985a,Sternberg_1995}. This approach may not be fully accurate. However, the state-specific rate coefficients of the reaction OH + H$_2$(v) are unknown except for v=1 \citep{Zellner_1981,Zhang_1994,Truong_1995}. This approximation gives coherent results with the state-specific rate coefficient determined in the latter studies. Moreover, when this approximation is not adopted and only the thermal rate coefficient is used, the abundance of H$_2$O is reduced by a factor 3. We expect the true value of the abundance to be between these two limits. Thus, this assumption is valid as a first approximation. Regarding the reaction O+H$_2$, we amended the version of the Meudon PDR code used in this paper to include the H$_2$~(v,J) state-specific rate coefficients recently computed by \cite{Veselinova_2021}. The H$_2$O and OH photodissociation rates are consistently computed by integrating the cross-section over the local radiation field, using the compiled cross section from the Leiden database \citep{Heays_2017}.

	\begin{table}[!h]
		\caption{Input parameters of the Meudon PDR Code.}      
		\label{table:inputs}      
		\centering                        
		\begin{tabular}{c c}       
			\hline\hline                 
			Parameters & Values  \\    
			\hline                      
			$A_V^{tot}$ & 20  \\   
			$P_{\rm th}/k$ (K cm$^{-3}$) & 10$^5$-10$^6$-10$^7$-\textbf{10$^8$}-10$^9$ \\
			$G_0^{\rm incident}$ (Mathis Unit) & 10$^2$-10$^3$-\textbf{10$^4$}-10$^5$  \\
			\hline 
			Transfer & FGK approximation \\
			Cosmic rays (s$^{-1}$ per H$_2$) & 5$\times$10$^{-17}$ \\
			R$_V$ & 3.1 \\
			$N_{\rm H}$/E(B-V) (cm$^{-2}$) & 5.8$\times$10$^{21}$ \\ 
			Dust to gas ratio & 0.01 \\
			Grain size distribution & $\propto \alpha^{-3.5}$ \\
			min grain radius (cm) & 1$\times$10$^{-7}$ \\
			max grain radius (cm) & 3$\times$10$^{-5}$ \\
			\hline                                   
		\end{tabular}
	\end{table}
	
	\subsection{Excitation of OH}
	\label{intensities_MIR}
	
	The OH excitation was computed in concert with the chemistry and the thermal balance using the method of \citet{GonzalezGarcia2008} that takes into account radiative pumping and collisional excitation, and allows to include the formation of species in excited states. In order to include the impact of OH production in rotationaly excited states, we assumed that only H$_2$O photodissociation in the $114-143~$nm UV range leads to the production of OH with a non-thermal state distribution and that the destruction pathways are not state-specific, that is, the destruction rate of an OH molecule is independent of its state. Any other formation route but H$_2$O photodissociation in the $114-143~$nm range is assumed to produce OH with a thermal state distribution at the local temperature of the gas. Therefore we neglected prompt emission of OH induced by water photodissociation at longer wavelength which produces vibrationally hot but rotationally cold OH \citep{vanHarrevelt2001}, and chemical pumping by O+H$_2$ which produces OH in lower rotational states $N<25$ \citep[][and A. Zanchet, priv. com.]{Liu_2000}. These excitation processes do not impact the highly rotationally excited lines of OH in the $9-15$ $\mu$m range that are the focus of the present study \citep[see discussion in][]{Tabone_2021}.
	
	These assumptions lead to the detailed balance equation
	\begin{equation}
		\sum_{j \neq i} P_{ji} n_j - n_i \sum_{j \neq i} P_{ij} +  F_{pd} \bar{f_{i}} + (F-F_{pd}) f_i(T_K)  - F \frac{n_i}{n(OH)}  = 0,
		\label{eq:statistical-eq}
	\end{equation}
	where $n_i$ $[\text{cm}^{-3}]$ is the local population densities of OH at a given position in the PDR. $P_{ij}$ are the radiative and collisional transition probabilities and include the contribution of the line and dust emission to the local radiation field \citep[see][for further details]{GonzalezGarcia2008}. We considered collisional (de-)excitation of OH with He and H$_2$ using collisional rate coefficients of \citet{Klos2007} and \citet{Offer1994} that have been further extrapolated to include collisional transitions between higher rotational levels of OH as in \citet{Tabone_2021}. $F$ is the total formation rate of OH and $f_i(T_K)$ is the Boltzmann distribution at a temperature $T_K$. $F_{pd}$ is the production rate of OH via H$_{2}$O photodissociation in the $114-143~$nm band $[\text{cm}^{-3} \text{s}^{-1}]$ as defined by:
	\begin{equation}
		F_{pd} = n_{H_2O} \int_{114~nm}^{143~nm} \sigma(\lambda) I(\lambda) d\lambda,
		\label{eq:Fpd}
	\end{equation}
	where $\sigma(\lambda)$ is the photodissociation cross section of H$_2$O, $I(\lambda)$ is the local UV radiation field, and $n_{H_2O}$ is the local number density of H$_2$O. Finally, $\bar{f_i}$ in Eq. (\ref{eq:statistical-eq}) is the state distribution of OH following H$_2$O photodissociation. The exact distribution $\bar{f_i}$ produced by water photodissociation in the $114-143~$nm UV range depends relatively weakly on the shape of the local radiation field and in this work, we adopt the distribution produced by Lyman-$\alpha$ photons ($\lambda=121.6~$nm) presented in \citet{Tabone_2021}. 
	
	The list of OH levels and radiative transitions stems from \citet{Tabone_2021} who used data from \citet{Yousefi2018} and \citet{Brooke2016}. In order to reduce the computational time, the number of OH levels have been reduced to a total of 412 by limiting the vibrational quantum number to $\varv \le 1$ and including only the OH(X) electronic ground state. All the rotational levels that are stable within a vibrational state are retained, which corresponds to $N \le 50$ and $N \le 48$ for $\varv = 0$ and $1$, respectively.  In order to account for prompt emission induced by the production of OH in the levels that have been discarded we use the reduced state distribution $\bar{f_i}$ derived by Tabone et al. (in prep.). Each rotational level is further split by the spin orbit coupling and the $\Lambda$-doubling. Following \citet{Tabone_2021}, we consider intra- and cross-ladder rotational transitions in the $\varv=0$ and $\varv=1$ bands as well as between the $\varv=1$ and $\varv=0$ states, resulting in a total of 2360 (ro-)vibrational transitions.

	\section{Results}
	
	First, we present the results for our fiducial model, that corresponds to a high pressure ($P_{\rm th}/k$ = 10$^8$ K cm$^{-3}$) and strong incident UV field model ($G_0^{\rm incident}$ = 10$^4$). This model is particularly interesting because these are representative parameters for a PDR for which OH mid-IR lines are the brightest. Then, we explore a grid of models to investigate how the line
	intensities vary with the pressure and incident UV field.
	
	\label{Results}
	\subsection{High pressure and UV field fiducial model} 
\label{ref_model}

\subsubsection{H$_2$O density profile and UV field}
\label{OH_H2O}
\begin{figure}
	\hfill\includegraphics[width=\linewidth]{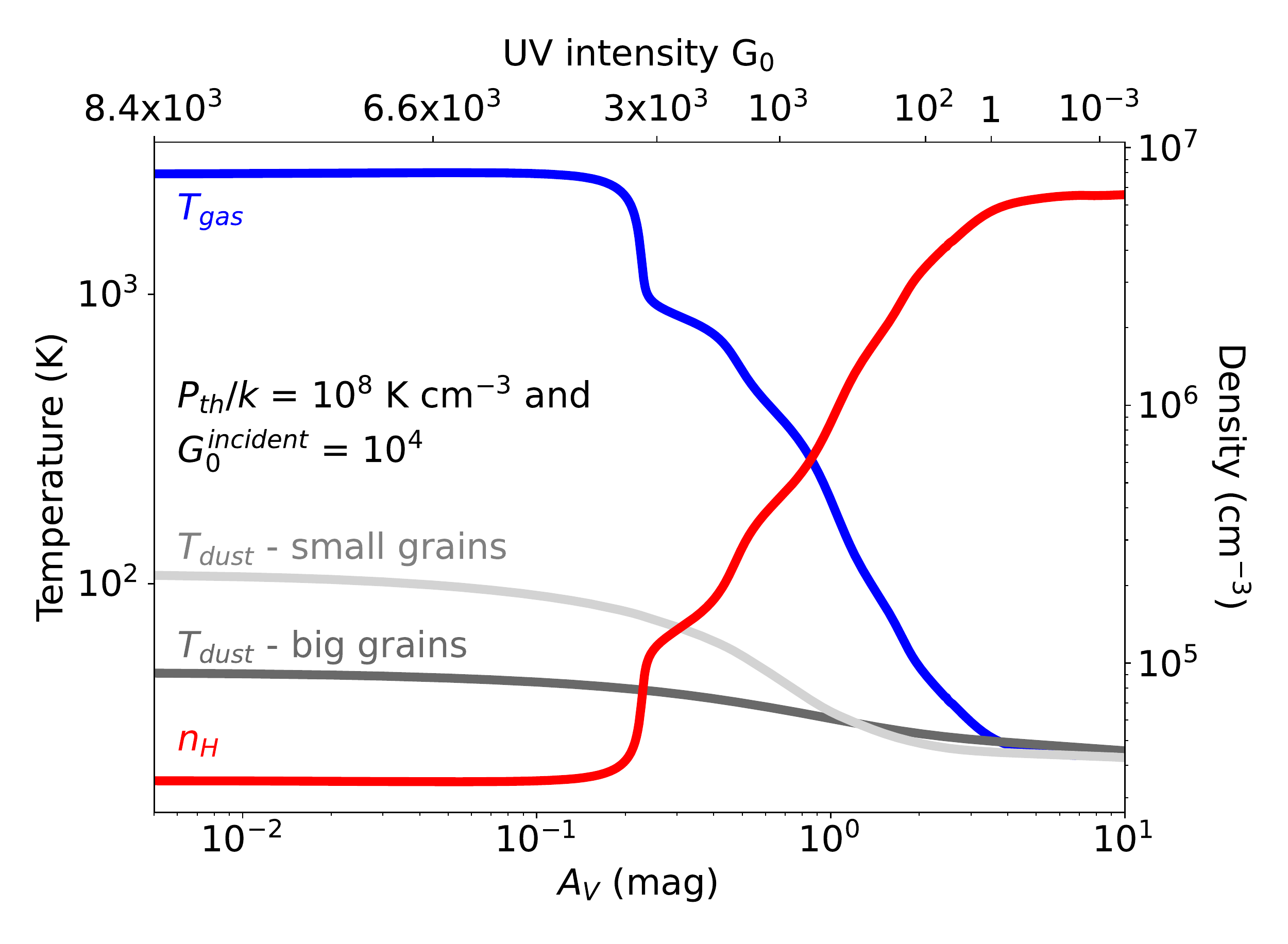}
	\includegraphics[width=\linewidth]{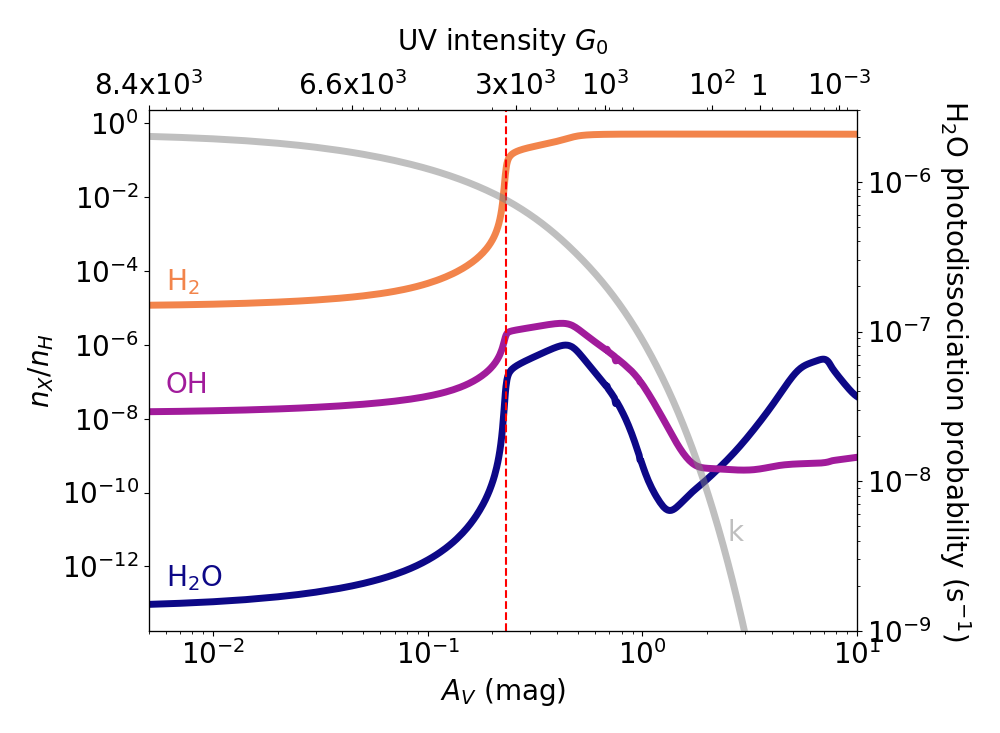}
	\caption{Results of our fiducial model with $P_{\rm th}/k$ = 10$^8$ K cm$^{-3}$ and $G_0^{\rm incident}$ = 10$^4$. (Top panel) Gas density, and gas and dust temperature as a function of visual extinction. The size of the small grains (resp. big grains) is 10$^{-7}$ cm (resp. 3.10$^{-5}$ cm). (Bottom panel) H$_2$O, OH and H$_2$ abundance as a function of visual extinction. H$_2$O photodissociation rate $k$ ($s^{-1}$) is also represented in light grey. The vertical red line shows the position of the H$^0$/H$_2$ transition. The local UV field intensity is given at 7 different positions on the top of the graph.}
	\label{fig:nH_T_nH2O_nOH}
\end{figure}

Figure \ref{fig:nH_T_nH2O_nOH} shows the total hydrogen density and gas temperature (top panel), and the H$_2$O, OH, and H$_2$ abundance (bottom panel) across the PDR. The H$_2$O photodissociation rate is also displayed in the bottom panel. Because thermal pressure is assumed to be constant across the PDR, the gas gets denser as the temperature drops with depth. Until $A_V$ = 0.1, 
the thermal balance is dominated by heating by the photoelectric effect, and [OI] and [CII] radiative cooling. 
Then, at the H$^0$/H$_2$ transition (driven by dust opacity and self-shielding), the temperature decreases steeply as the gas is primarily cooled down by H$_2$ emission and as the heating by photoelectric effect is less efficient due to the attenuation of the FUV field. Deeper into the PDR ($A_V$ $>$ 1), the gas temperature smoothly decreases due to the cooling by CO, and eventually by gas-grain thermal coupling \citep{Tielens_1985a}.

\begin{figure}
	\centering
	\includegraphics[width=\linewidth]{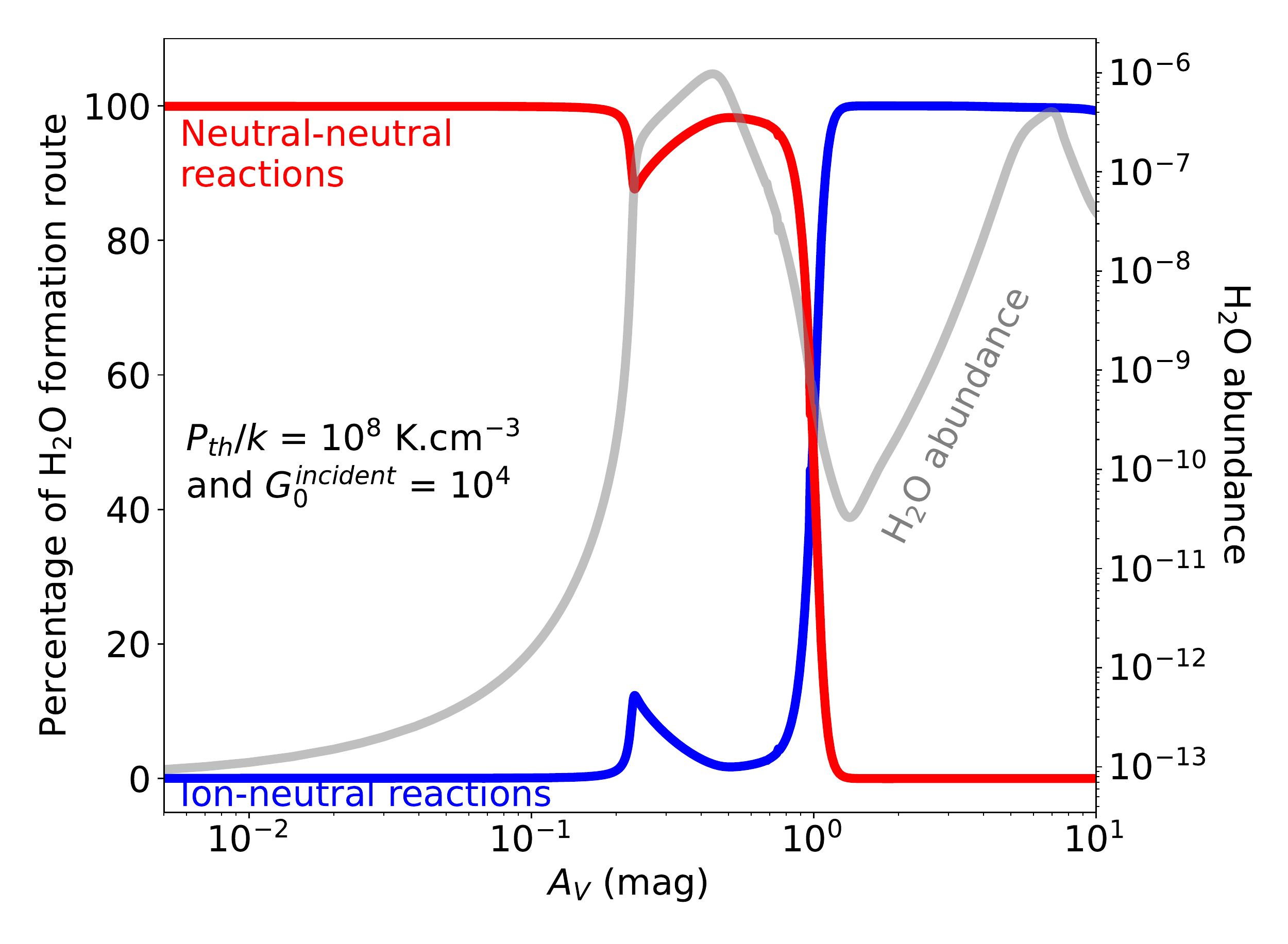}
	\caption{Percentage of the different formation routes of H$_2$O as a function of the visual extinction for the model $P_{\rm th}/k$ = 10$^8$ K cm$^{-3}$ and $G_0^{\rm incident}$ = 10$^4$. The blue line represents the percentage of the ion-neutral formation route, and the red line represents the percentage of the neutral-neutral formation route that requires high temperatures ($T_{\rm K} \gtrsim 300~$K). The increase in the ion-neutral formation route around $A_V$ = 0.2 is due to the increase in OH$^+$ following neutral-neutral formation of OH. It is therefore associated to warm chemistry.}
	\label{fig:H2O_chemistry}
\end{figure}

The H$_2$O and OH abundance profile reflects that of H$_2$ and of the temperature. Two peaks in H$_2$O abundance, corresponding to two distinct reservoirs of water are highlighted in the bottom panel of Fig. \ref{fig:nH_T_nH2O_nOH}. The first peak at $A_V$ $\sim$ 0.5 in the warm and irradiated region is mainly due to neutral-neutral reactions (see Fig. \ref{fig:H2O_chemistry}). Here, H$_2$O forms primarily via a two step process with energy barriers ($\Delta$E) \citep[][see \ref{neutral_reaction}]{Van_Dishoeck_2013}: \begin{align}
	\label{OH_form}
	&\ce{O + H$_2$ <=> OH + H} \qquad \Delta \text{E = +2401 K.} \\
	\label{H2O_form}
	& \ce{OH + H$_2$ <=> H$_2$O + H} \qquad \Delta \text{E = +1751 K.}
\end{align}
Since it corresponds to an irradiated reservoir of H$_2$O, OH prompt emission is expected to be confined to that layer. The position of the first H$_2$O abundance peak is a compromise between being deep enough in the cloud to have the presence of molecular H$_2$ (not photodissociated) necessary to the formation of water and being close to the edge to have a temperature high enough to overcome the energy barrier of the chemical reaction. The peak is indeed located close to the H$^0$/H$_2$ transition, in agreement with the study of \cite{Sternberg_1995} (see their Figs. 8 and 9). Therefore, the peak abundance of H$_2$O is highly sensitive to the temperature, as further shown in Sect. \ref{results_grid} and in appendix \ref{summary_appendix}. Interestingly, in the warm region of interest, the inclusion of the enhanced reactivity of excited H$_2$ in the formation of OH (and therefore H$_2$O) does not change drastically the results compared with the thermal rates. For example, at the H$_2$O abundance peak, the formation rate of H$_2$+O is only increased by a factor 2 compared to the thermal rate which is rather low for these order of magnitude. This is related to the relatively modest endoergicity of reaction O + H$_2$ compared to hydrogenation reactions of N and S$^+$ \citep[e.g.][]{Goicoechea_2022}. This lower endoergicity results in relatively high rate coefficients for the lower-v H$_2$ states. In the warm region, OH is also primarily produced by neutral-neutral reactions (see Eq. \ref{OH_form} and appendix \ref{appendix_chemistry_OH}) with a significant contribution of H$_2$O photodissociation at the peak of OH abundance. However, as explained further (see Sec. \ref{OH_midIR_prediction}), only water photodissociation can lead to OH excited in highly rotational states. Thus, OH mid-IR emission depends on water abundance and not directly on OH abundance.

Deeper into the PDR ($A_V>0.5$), the H$_2$O and OH abundances decrease drastically because the temperature is dropping and both molecules are still efficiently destroyed by photodissociation (see the H$_2$O photodissociation rate in Fig. \ref{fig:nH_T_nH2O_nOH}, bottom panel). The second water reservoir peaks deeper into the cloud in colder regions, with a peak in H$_2$O abundance at $A_V \simeq 7$. In this region, H$_2$O is primarily formed via ion-neutral reactions, ending with the electronic recombination of H$_3$O$^+$:
\begin{equation}
	\ce{OH^+ ->[H_2] H_2O^+ ->[H_2]   H_3O^+ ->[e^-] H_2O},    
\end{equation}
OH$^+$ being formed by O + H$_3^+$ in molecular regions (see Fig. \ref{fig:H2O_chemistry} and
appendix \ref{H2O_abundance_2}). Water is still primarily destroyed by photodissociation with an efficiency that decreases with depth into the PDR (see light grey in Fig. \ref{fig:nH_T_nH2O_nOH}). Thus the H$_2$O abundance increases again with $A_V$. We note that deep into the cloud, H$_2$O formation on grains followed by desorption (not included in our model) is expected to be relevant \citep{Hollenbach_2009,Putaud_2019}. However, this cold component is not of interest in this study since it produces a negligible amount of highly excited OH. Indeed, although the cold reservoir is in larger proportion than the warm reservoir, it is so weakly irradiated that the amount of water photodissociated coming from this region is negligible in comparison to the one coming from the warm region.

Interestingly, the amount of warm H$_2$O (N(H$_2$O) $\sim$ 4$\times$10$^{14}$ cm$^{-2}$) represents only 5-15\% of the total H$_2$O content of the PDR. This is in agreement with the results of \cite{Putaud_2019} that suggest that the cold component dominates the emission of H$_2$O in \textit{Herschel} data. This explains why we cannot use these previous results to predict OH mid-IR emission as we are only interested in the warm reservoir.

\subsubsection{Prediction of OH mid-IR lines}
\label{OH_midIR_prediction}

\begin{figure}[b]
	\centering
	\includegraphics[width=\linewidth]{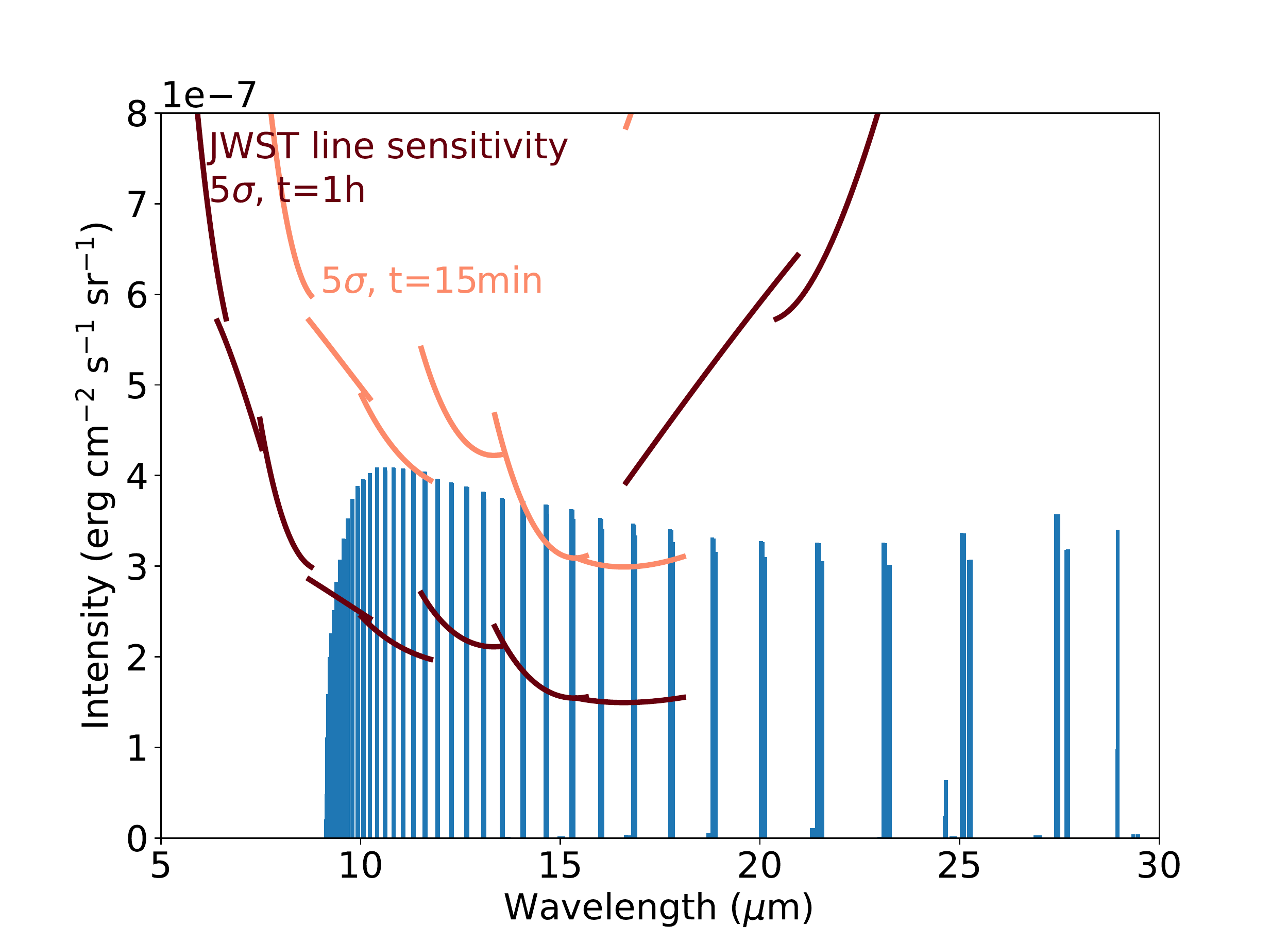}
	
	\caption{Intensities of OH mid-IR lines observed with a viewing angle of 60$\degree$ as a function of wavelength for a model $P_{\rm th}/k$ = 10$^8$ K cm$^{-3}$ and $G_0^{\rm incident}$ = 10$^4$. The brown and orange lines represent the \textit{JWST} sensitivity for corresponding integration time and an SNR of 5 \citep{Glasse_2015}.}
	\label{fig:spectre_OH}
\end{figure}

\begin{figure}
	\centering
	\includegraphics[width=0.7\linewidth]{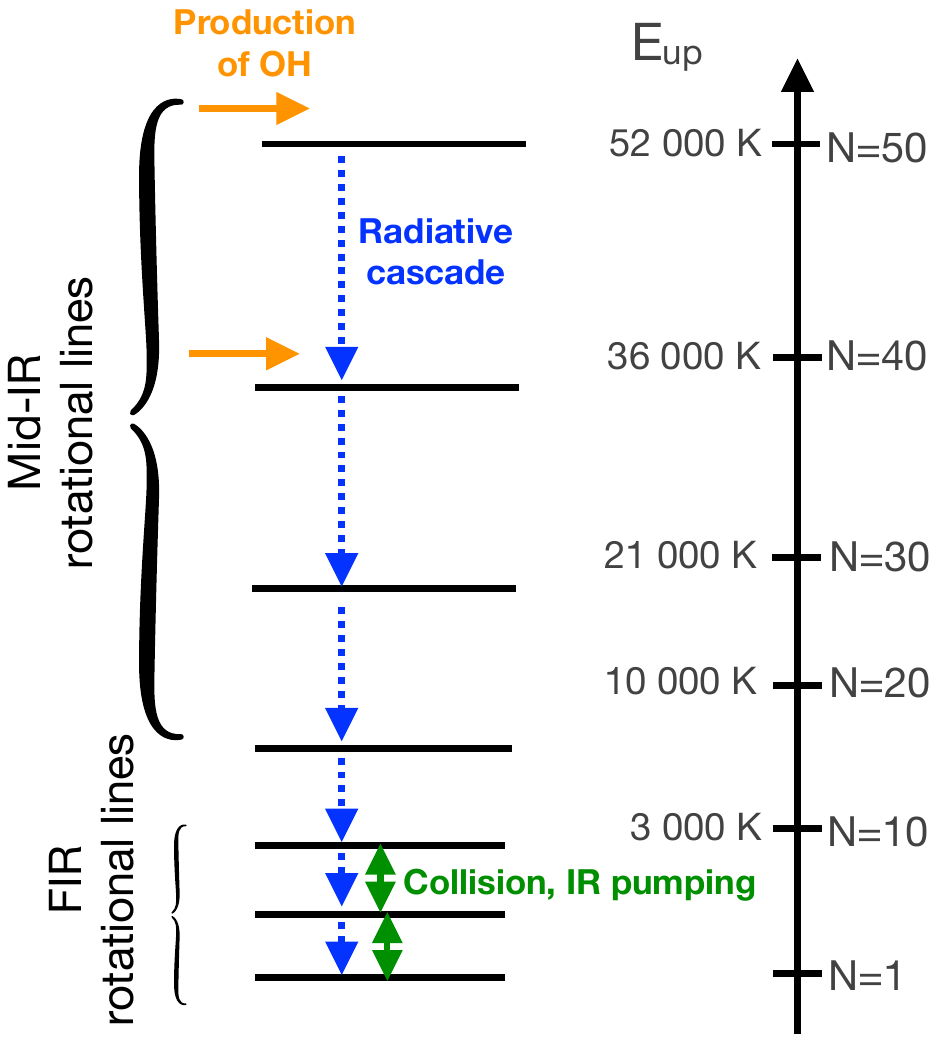}
	\caption{Schematic of the excitation of OH in the presence of prompt emission. The production of rotationnally excited OH (orange arrows) is followed by a $N\rightarrow N-1$ radiative cascade  (pictured by blue arrows), emitting the series of lines shown in Fig. \ref{fig:spectre_OH}. For lower rotational states ($N\lesssim 10$), OH levels can be excited by IR pumping and collisions which emitting in the far infrared. In this greatly simplified schematic, only a limited number of rotational levels are shown, ignoring $\Lambda$-doubling and spin-orbit coupling split of rotational levels, vibrational levels, and other kinds of weaker radiative transitions.}
	\label{fig:schematics_OH_exc}
\end{figure}

Figure \ref{fig:spectre_OH} presents the calculated mid-IR spectrum of OH for our fiducial model with a viewing angle of 60$\degree$. The viewing angle is the angle between the line of sight and the normal to the PDR, with 0$\degree$ being face-on and 90$\degree$ edge-on. The value of 60$\degree$ is representative of the inclination of most of the observed interstellar PDRs as they are closer to being edge-on than face-on (e.g. the Orion Bar, the Horsehead Nebula, ...). A series of pure-rotational lines can be seen in the 9-27 $\mu$m range coming from high-N states (15 $\leq$ N $\leq$ 45) with upper energies > 5000 K. These pure intra-ladder rotational lines, that are split in four components by $\Lambda$-doubling and spin-orbit coupling (not apparent in Fig. \ref{fig:spectre_OH}), dominate the mid-IR OH spectrum. There is a steep increase in line intensity with decreasing N shortward of $\lambda=$10 $\mu$m (N$\simeq$35) and then a slow decrease longward of $\lambda=$10 $\mu$m.  

Figure \ref{fig:schematics_OH_exc} summarizes the excitation process leading to the observed prompt emission spectrum shown in Fig. \ref{fig:spectre_OH}. IR radiative pumping cannot excite these lines due to the high energy of the upper levels. De-excitation by stimulated emission by the IR background is also negligible because of the small photon occupation number. Collisional excitation  can not populate the upper levels due to their very high upper energy level, and collisional desexcitation is negligible due to the very high critical densities of these levels ($n_{\rm crit} \geq$ 10$^{13}$ cm$^{-3}$). Hence, the population of the levels are only set by the radiative cascade following H$_2$O photodissociation forming OH in high-N states. 

In particular, we recover the result of \citet{Tabone_2021} that the OH mid-IR lines intensities are only proportional to the column density of water photodissociated in the $\tilde{B}$ band (114 nm $< \lambda <$ 143nm)
\begin{equation}
	\Phi_{B} = \int_z F_{pd}(z) dz,
\end{equation}
where we recall that $F_{pd}$ is the (volumic) destruction rate of H$_2$O via photodissociation in the $114-143~$nm UV range (see Eq. ($\ref{eq:Fpd}$). This simple result demonstrates that OH mid-IR lines give a direct access to H$_2$O photodissociation but no (direct) information on the column density of OH, the density, or the temperature. H$_2$O abundance in irradiated media being highly sensitive to the physical conditions (in particular temperature), OH mid-IR lines are in turn an indirect but very sensitive diagnostics of the physical conditions.

We also recover that the overall shape of the mid-IR spectrum of OH depends neither on OH column density nor on the photodissociated column density of H$_2$O $\Phi_{B}$. In fact, as discussed in \citet{Tabone_2021}, the relative intensity of the excited lines is only set by the distribution of nascent OH, which is set by the spectral shape of the UV field. However, in this work, we neglect the effect of the shape of the radiation field within the $114-143~$nm range and we take the distribution of OH at Ly-$\alpha$ wavelength ($121~$nm) as a representative wavelength. Therefore, the intensity of each line shown here is, by construction, only proportional to $\Phi_{B}$ and does not depend on other parameters. In the following, we focus our study on the total intensity of the $N=30 \rightarrow 29$ quadruplet at 10.8 $\mu$m as it is the brightest of the spectrum.

\begin{figure}[!b]
	\centering
	\includegraphics[width=\linewidth]{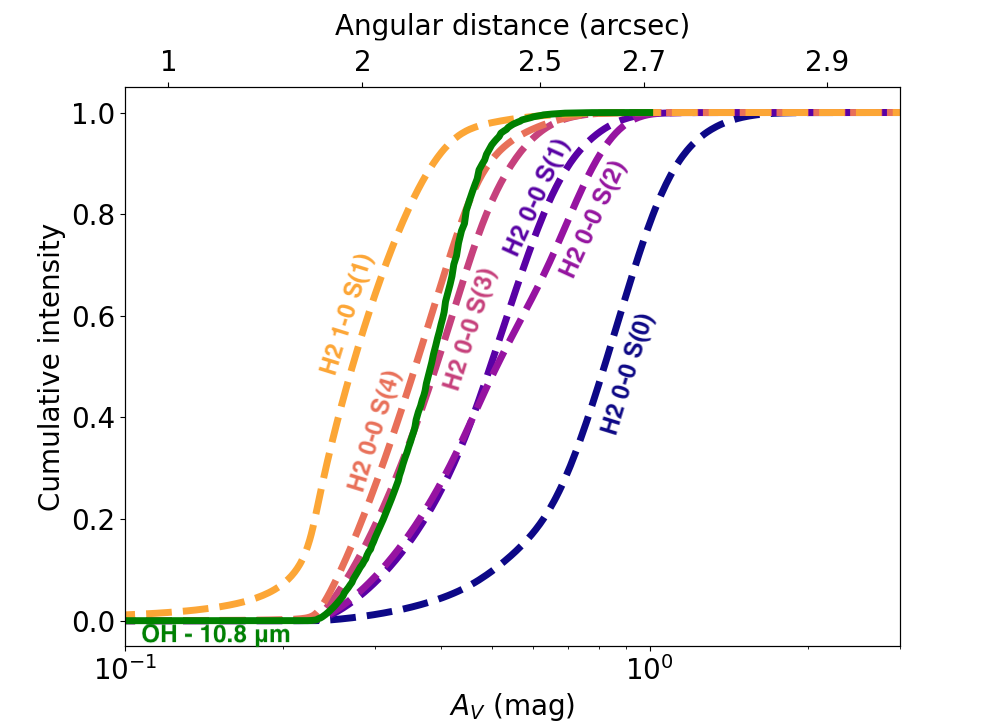}
	\caption{Cumulative intensity (normalized) of the 10.8 $\mu$m OH line, and of several rotationnal and ro-vibrational lines of H$_2$ calculated with the Meudon PDR code with a thermal pressure $P_{\rm th}/k$ = 10$^8$ K cm$^{-3}$ et $G_0^{\rm incident}$ = 10$^4$. OH mid-IR emission traces a thin layer close to the H$^0$/H$_2$ transition.}
	\label{fig:OH_H2_position}
\end{figure}

In order to locate the emitting region of rotationally excited OH emission in the PDR, we calculate the cumulative intensity\footnote{$I$($A_V$)/$I$($A_V$=20) assuming OH and H$_2$ lines are optically thin.} as function of the PDR depth. Fig. \ref{fig:OH_H2_position} compares the cumulative intensity of 10.8 $\mu$m OH line and rotationally excited H$_2$ lines across the PDR. First, this figure shows that OH emission is confined to a thin layer around $A_V$ $\simeq$ 0.4 that corresponds to the H$^0$/H$_2$ transition. It can also be seen that this OH line peaks at the same position as the relatively excited H$_2$ pure-rotational lines such as 0-0 S(3) and 0-0 S(4). H$_2$ rotational lines are excited by collision so their emitting region reflect the temperature gradients across the PDRs. Thus, 0-0 S(3) ($E_{\rm u}/k$ $\sim$ 2504 K) and 0-0 S(4) ($E_{\rm u}/k$ $\sim$ 3474 K) peak closer to the edge than less excited lines such as 0-0 S(0) ($E_{\rm u}/k$ $\sim$ 510 K), which has a lower upper energy level. The H$_2$ 1-0 S(1) line ($E_{\rm u}/k$ $\sim$ 6952 K) is mainly populated by UV pumping. This explains why this line emission peaks closer to the edge at the H$^0$/H$_2$ transition. 

We note that H$_2$ rotational lines is a good diagnostic of the gas temperature as they are close to local thermodynamical equilibrium and optically thin. However, this diagnostic needs high angular resolution to separate the emitting region of each lines, otherwise, the line fluxes are averaged on the PDR. 
The gas temperature profile  can therefore only be measured with the H$_2$ lines in close and not too narrow PDRs. In the cases where the observations are not spatially resolved, OH mid-IR lines can be a unique indirect diagnostic to probe temperature at the H$^0$/H$_2$ transition as OH mid-IR emission is strongly correlated to the thermal balance due the formation of H$_2$O needing high temperatures.

\subsection{Effects of thermal pressure and UV field}
\label{results_grid}

As shown in Sect. \ref{ref_model}, OH mid-IR emission is directly proportional to the amount of photodissociated water, water abundance being very sensitive to the temperature and the local UV field. Therefore, one would naturally expect a strong dependency of the OH line intensity on the thermal pressure and the strength of the incident UV radiation field. In this section, we study a grid of models with pressure ranging from $P_{\rm th}/k$ = 10$^5$ to 10$^9$ K cm$^{-3}$ and incident UV field intensity $G_0^{\rm incident}$ = 10$^2$ to 10$^5$.

\begin{figure}
	\centering
	\includegraphics[width=\linewidth]{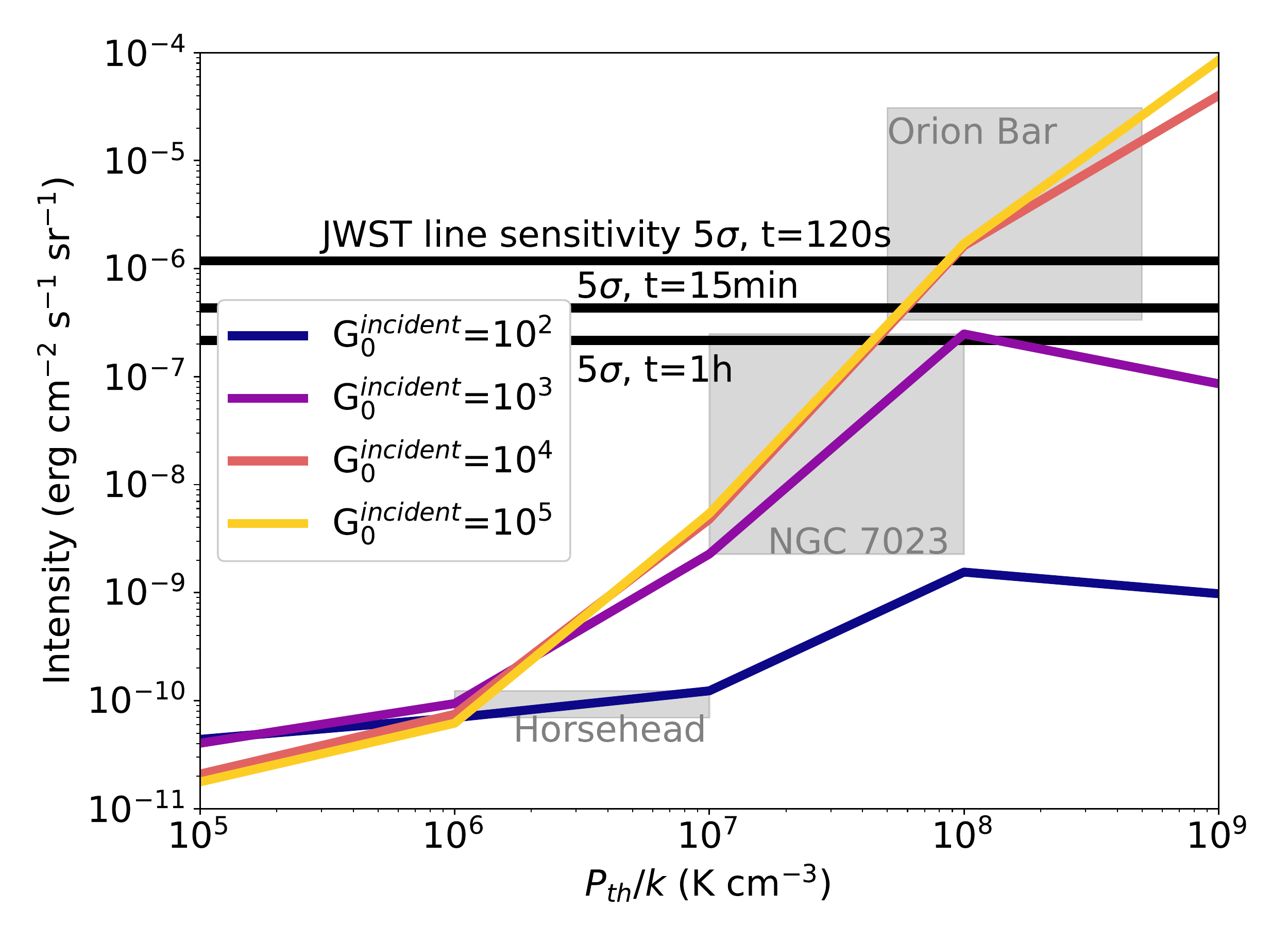}
	\caption{Summed intensity of the quadruplet at $\lambda$ = 10.8 $\mu$m observed with a viewing angle of 60$\degree$ as a function of the gas pressure for different UV field intensities. The black horizontal lines represent the \textit{JWST} sensitivity for corresponding integration time and an SNR of 5. 
	}
	\label{fig:I30_Pgas_total}
\end{figure}

\begin{figure*}
	\centering 
	\includegraphics[width=0.49\linewidth]{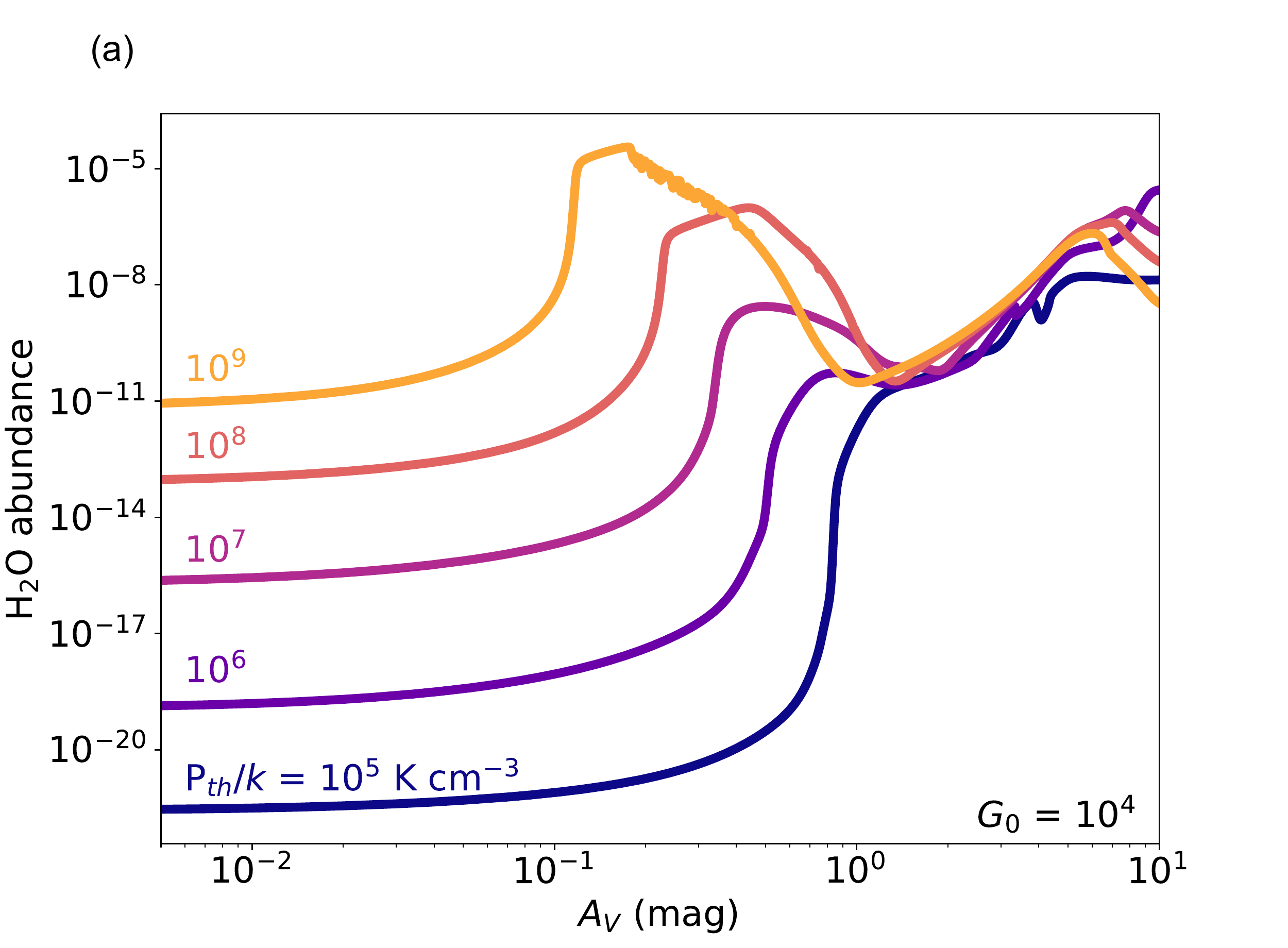} \includegraphics[width=0.49\linewidth]{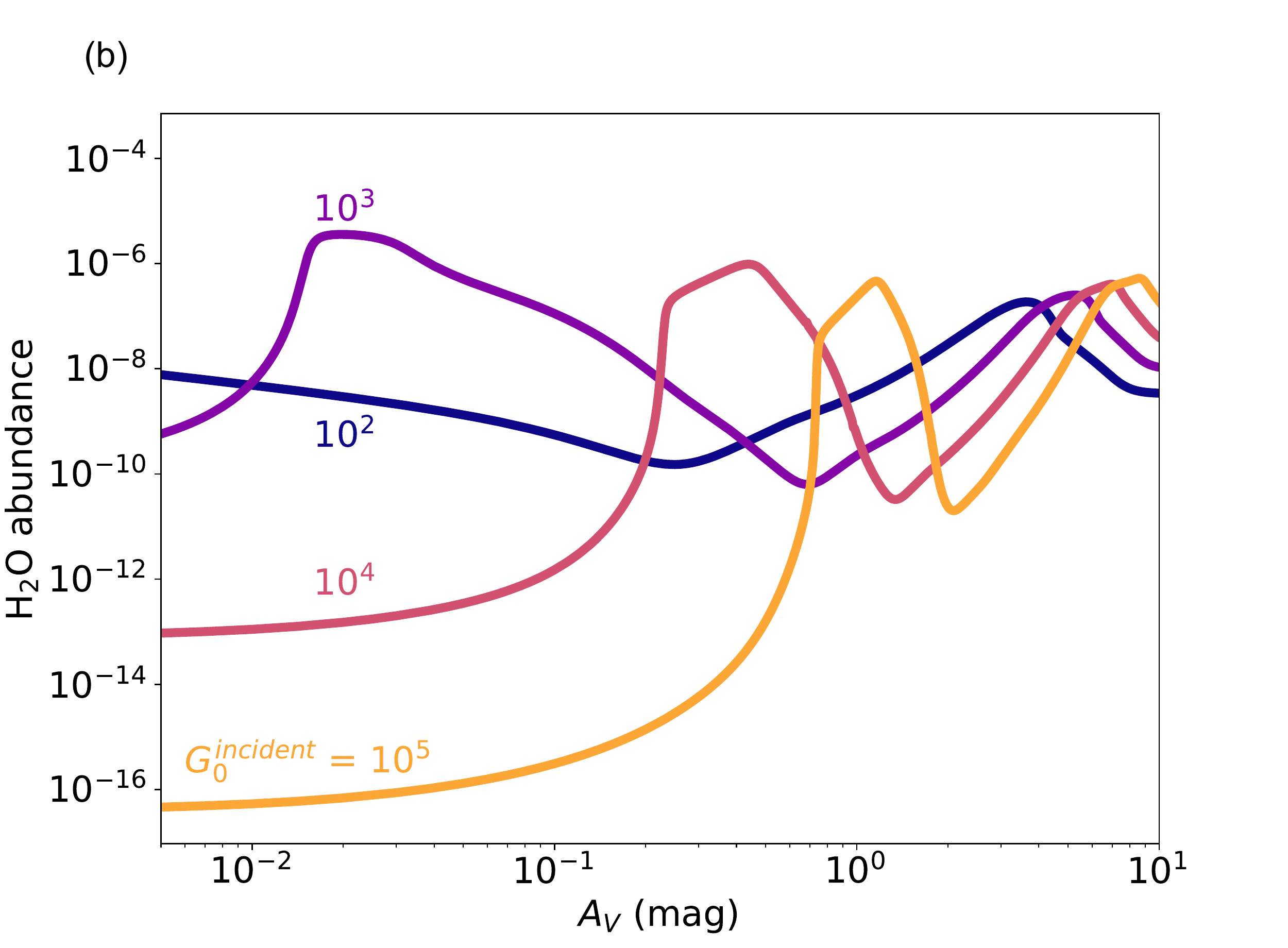} \\
	
	\includegraphics[width=0.49\linewidth]{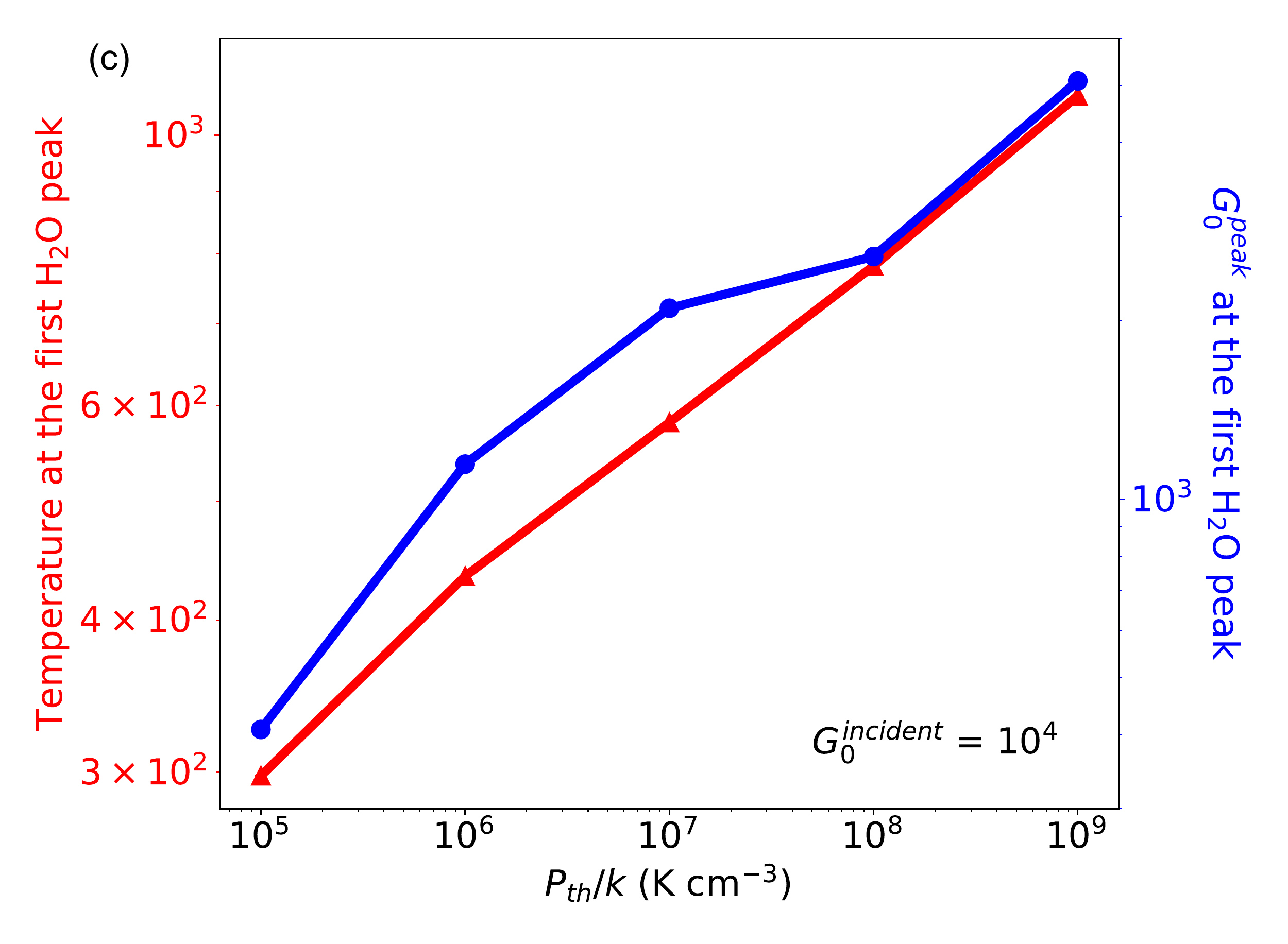} \includegraphics[width=0.49\linewidth]{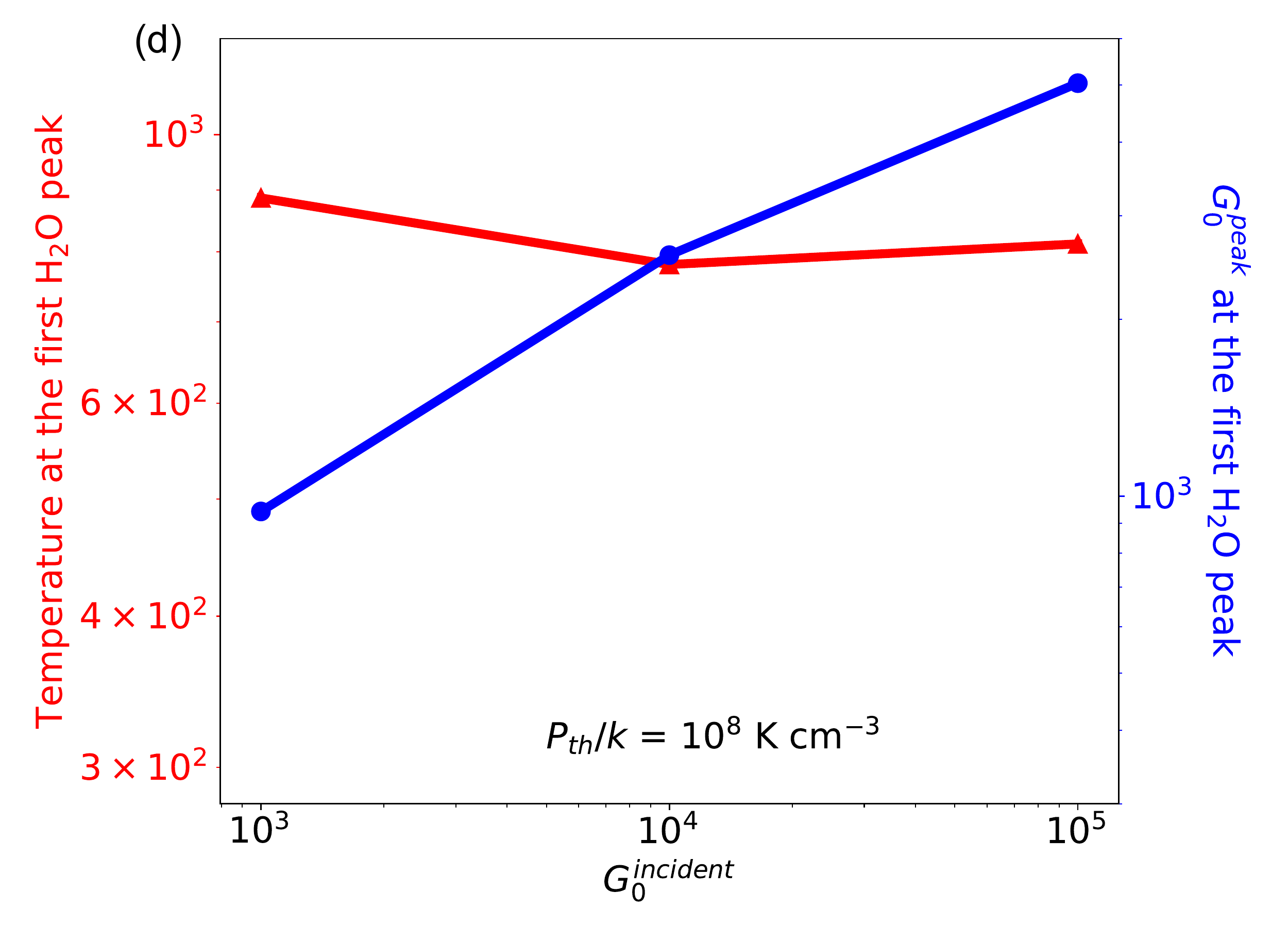}
	\caption{Evolution of H$_2$O abundance, local gas temperature and local UV field as a function of the thermal pressure and the incident UV field. (Top panels) Abundance profile of H$_2$O for different thermal pressures at a given UV field intensity (a) and for different incident UV fields at a given thermal pressure (b). (Bottom panels) Evolution of the temperature and the UV field at the first H$_2$O peak as a function of the thermal pressure (c) and the incident UV field intensity (d).}
	\label{fig:nH2O_var}
\end{figure*}

Figure \ref{fig:I30_Pgas_total} summarizes the evolution of the 10.8 $\mu$m line intensity as a function of the thermal pressure for different UV field intensities. It can be seen that OH emission depends strongly on the thermal pressure (see Sect. \ref{pressure} for further explanation). On the other hand, for UV field intensity $G_0^{\rm incident} >$ 10$^3$, OH line intensities do not depend much on incident UV field intensity (see Sect. \ref{UV_field} for further explanation). To understand this result, we study the evolution of the temperature and UV field intensity at the warm peak of H$_2$O as a function of the thermal pressure and the intensity of the UV field $G_0^{\rm incident}$ as presented in Fig. \ref{fig:nH2O_var}. 

\subsubsection{Dependence on pressure}
\label{pressure}
Figure \ref{fig:nH2O_var}-a displays the abundance profile of H$_2$O for different thermal pressures. It shows that we still recover a peak in H$_2$O abundance down to $P_{\rm th}/k$ = 10$^6$ K cm$^{-3}$, corresponding to the warm and irradiated reservoir of H$_2$O that produces rotationally excited OH. As thermal pressure increases, the warm H$_2$O reservoir moves closer to the edge and the corresponding peak abundance of H$_2$O increases dramatically by 6 orders of magnitude from $P_{\rm th}/k = 10^6$ to $10^9$ K cm$^{-3}$. This result might be surprising since at high pressure, the warm reservoir is more irradiated (see Fig. \ref{fig:nH2O_var}-c, blue curve) and therefore H$_2$O is more efficiently photodestroyed. The density also increases with pressure, which directly enhances the formation rate of H$_2$O in proportion. However, this effect is not enough to account for the dramatic increase in H$_2$O abundance. In fact, as pressure increases, the gas at the H$^0$/H$_2$ transition gets warmer (see Fig. \ref{fig:nH2O_var}-c, red curve), triggering active OH and H$_2$O formation via neutral-neutral reactions. Therefore, the steep increase in H$_2$O abundance with $P_{\rm th}/k$ is primarily due to the gas temperature rising. We still notice that the temperature increasing is an indirect consequence of the increase in density and local UV field at the H$^0$/H$_2$ transition since the H$_2$ forms closer to the PDR edge at high densities, and heating by H$_2$ UV pumping and H$_2$ formation are enhanced.

\begin{figure}[!h]
	\centering
	\includegraphics[width=\linewidth]{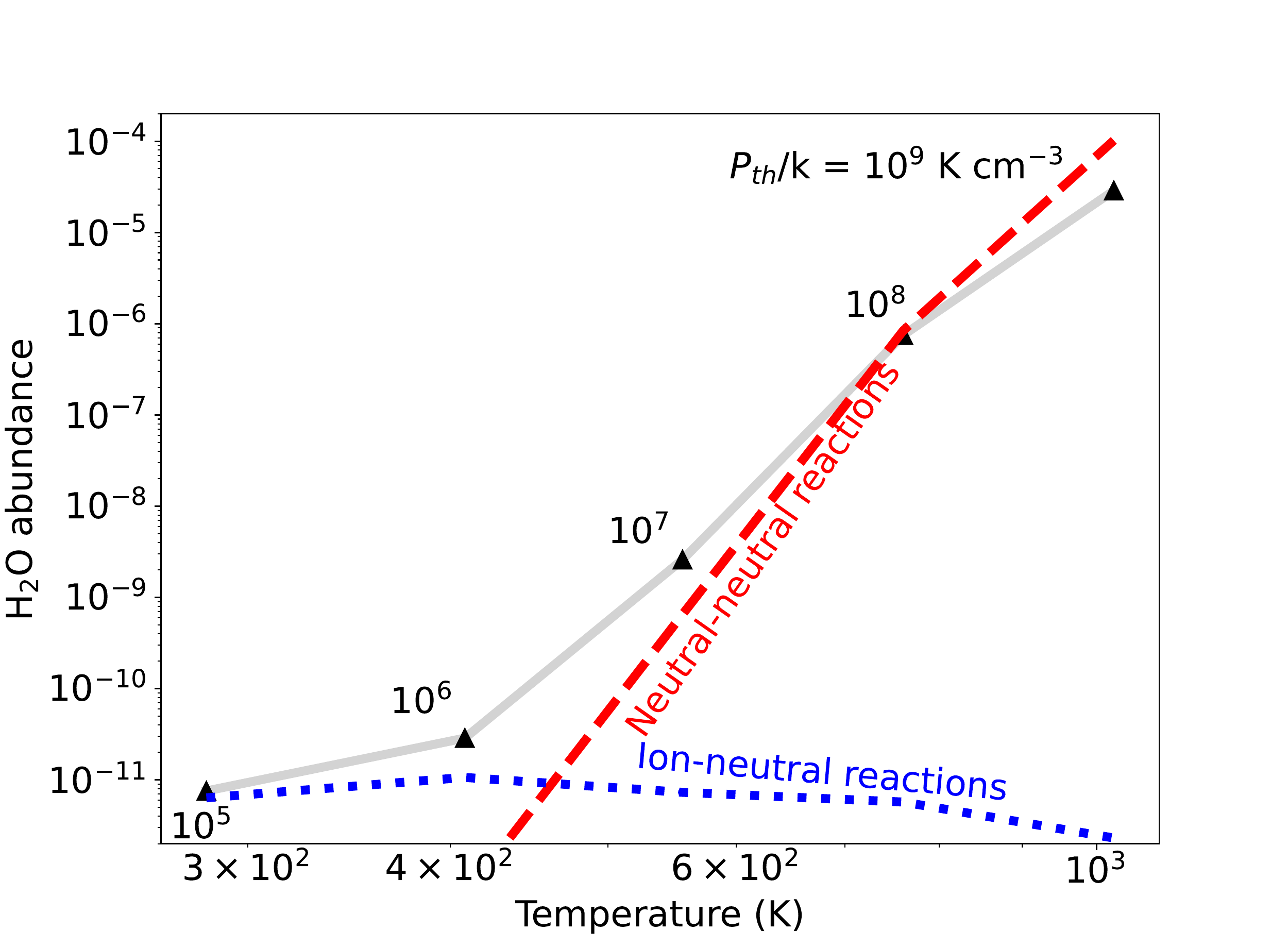}
	\caption{Analytic calculations of first H$_2$O peak abundance as a function of the temperature at the first H$_2$O peak \citep[appendix \ref{appendix_chemistry}, using thermal rate coefficient from][]{Agundez_2010,Veselinova_2021}. The state-specific rate coefficients are not considered in the analytic calculations. Using the state-specific chemistry enhance the abundance of H$_2$O by a factor 2. The dotted blue line represents the calculation of the abundance of H$_2$O formed by ion-neutral reaction and the dashed red line represents the calculation of abundance of H$_2$O formed by neutral-neutral reactions. The triangles are the first H$_2$O peak abundance calculated with the Meudon PDR code for models at incident UV field $G_0^{\rm incident}$ = 10$^4$ and different thermal pressure. 
	}
	\label{fig:H2O_chemistry_evolution}
\end{figure}

The sensitivity of the formation route of H$_2$O on temperature is further highlighted in Fig. \ref{fig:H2O_chemistry_evolution} where the calculated abundance of H$_2$O is compared to our analytic model of oxygen chemistry detailed in Appendix \ref{appendix_chemistry}. We recover the fact that down to $P_{\rm th}/k \simeq 10^7$ K cm$^{-3}$, H$_2$O is primarily formed by neutral-neutral reactions, and that the efficiency of this route declines for lower thermal pressures due to lower temperatures. Interestingly, at very low pressure, below $P_{\rm th}/k \simeq 10^6$ K cm$^{-3}$, the ion-neutral reaction, that is weakly dependent on temperature, takes over from the neutral-neutral route. This roughly corresponds to the thermal pressure below which the peak in warm H$_2$O abundance disappears.

We recall that mid-IR OH line intensities are proportional to the quantity of H$_2$O photodissociated which is the product of the H$_2$O density profile and the UV field flux integrated over the cloud. 
This explains why OH mid-IR lines are found to increase with thermal pressure (see Fig. \ref{fig:I30_Pgas_total}): H$_2$O is more efficiently formed via neutral-neutral reaction in the warm molecular layer and, to a lesser extent, because that reservoir is more irradiated (see Fig. \ref{fig:H2O_chemistry_evolution}).

\subsubsection{Dependence on incident UV field}
\label{UV_field}

Figure \ref{fig:nH2O_var}-b displays the H$_2$O abundance profile for different strengths of incident UV field and $P_{\rm}/k=10^8$ K cm$^{-3}$. As the strength of the incident UV field increases, the warm H$_2$O reservoir moves deeper into the cloud. The H$^0$/H$_2$ transition is indeed shifted to larger A$_V$ because the total column density needs to be higher to trigger the H$^0$/H$_2$ transition. The H$_2$O peak abundance is also somewhat reduced for stronger incident UV fields. This is due to both, a slight decline of the temperature that quenches H$_2$O formation by the neutral route, and to an increase in the local radiation field at the H$^0$/H$_2$ transition (see Fig. \ref{fig:nH2O_var}-d).  

With these results, the conclusion is that mid-IR OH line intensities, that are proportional to the amount of water photodestroyed, depend weakly on the incident UV field intensity for $G_0^{\rm incident} \gtrsim 10^3$. Indeed, as the incident UV field increases, the amount of H$_2$O decreases slowly but the local radiation field increases accordingly. Both effects tend to act against each other for OH mid-IR emission, resulting in a relatively weaker dependency on the incident UV field than on thermal pressure, at least for $G_0^{\rm incident} > 10^3$, where the H$_2$O abundance seems to saturate with increasing incident UV field.

\subsubsection{Evolution of the OH/H$_2$ lines ratio}

\begin{figure}
	\centering
	\includegraphics[width=\linewidth]{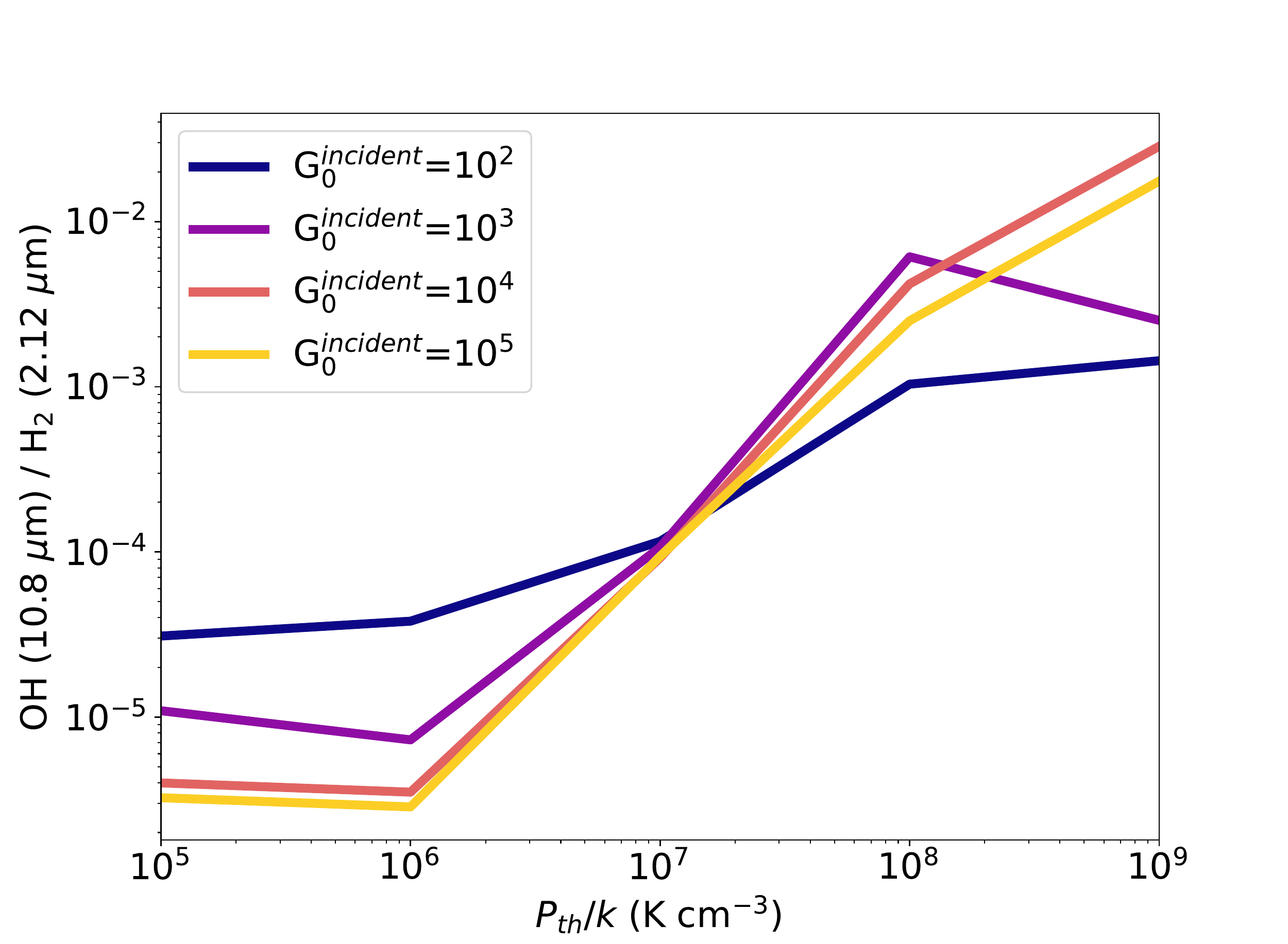}
	\caption{Ratio of the predicted OH line at 10.8 $\mu$m over the H$_2$ line 1-0 S(1) at 2.12 $\mu$m predicted by the Meudon PDR Code.}
	\label{fig:ratio_OH_line}
\end{figure}

The predicted line intensities depend on the inclination of the PDR which is a major source of uncertainty in observations. Therefore, we present in Fig. \ref{fig:ratio_OH_line} a more robust prediction, that is the ratio between the 10.8 $\mu$m OH line and the 2.12 $\mu$m H$_2$ ro-vibrational line, which peaks at a position close to that of OH and has already been observed at high angular resolution \citep[e.g.,][]{Habart_2022}.

Fig. 8 shows that the ratio is not constant and that over $P_{\rm th}/k\ge$ 
10$^6$ K cm$^{-3}$, the ratio increases. The higher the pressure is, the higher the warm gas quantity. This fosters warm H$_2$O formation by neutral-neutral reaction and thus fosters the formation of OH which emits in the mid-IR. The H$_2$ 1-0 S(1) line increases with pressure as it is proportional to the gas density \citep[for $n_H/G_0 < 40$ cm$^{-3}$ e.g.,][]{Burton90} but it is not dependent on the gas temperature.

\section{Application to the Orion Bar and other environments}
\label{Orion_bar}
\subsection{Orion Bar}
The Early Release Science program "PDRs4All: Radiative
feedback from massive stars" \citep{ERS_2022} for the \textit{JWST} observations is dedicated to studying the interactions of massive stars with their surroundings. The target of this program is a well-known PDR: the Orion Bar. This region will be observed through NIRSpec, NIRCam and MIRI, giving IFU spectroscopy with NIRSpec and MIRI, and imaging with NIRCam and MIRI. In this section, we use predictions for the Orion Bar to illustrate the potential of OH and discuss the main limitations of upcoming \textit{JWST} observations of interstellar PDRs.

\subsubsection{Predicted OH line intensities}
\label{OH_lines_OB}
The parameters used in the models of the Orion Bar are summarized in Table \ref{table:inputs_orionbar}. We consider an isobaric model with a pressure ranging from $P_{\rm th}/k$ = 5$\times$10$^7$ K cm$^{-3}$ to $P_{\rm th}/k$ = 5$\times$10$^8$ K cm$^{-3}$, in agreement with previous studies \citep{Allers_2005,Joblin_2018}. We adopt an incident UV field coming from an illuminating O7 star with an effective temperature $T_{\rm eff}$ = 40,000 K modeled by a blackbody at $T_{\rm eff}$. 

The UV field intensity is taken equal to 2.10$^4$ in Mathis units which is in agreement with previous estimates giving $G_0^{\rm incident}$ = 1-4$\times$10$^4$ \citep{Tielens_1985b,Marconi_1998}. 
We assume the extinction curve HD 38087 of \cite{Fitzpatrick_1990} and R$_V$ = 5.62 which is close to the value determined for the Orion Bar of 5.5 \citep{Marconi_1998}. This extinction curve is also in agreement with the recent dust study by \cite{Schirmer_2022} from \texttt{THEMIS} dust model in the Orion Bar with nano-grains depletion.

These models include an exact radiative transfer calculation for the UV pumping of H$_2$ lines originating from the first 30 levels of H$_2$, while the other lines are treated using the FGK approximation \citep{Federman_1979}. This allows to account for mutual shielding effects between overlapping H$_2$ and H UV absorption lines. This approximation can affect the position of the H$^0$/H$_2$ transition as the FGK approximation tends to shift the H$^0$/H$_2$ transition closer to the edge of the cloud, slightly affecting the abundance of H$_2$O in the warm layer. However, the emerging line intensities of OH and H$_2$ are little affected (less than 10\%).

\begin{table}[!h]
	\caption{Input parameters of the Meudon PDR Code for the Orion Bar Model.}      
	\label{table:inputs_orionbar}      
	\centering                        
	\begin{tabular}{c c}       
		\hline\hline                 
		
		Parameters & Values  \\    
		\hline                      
		$A_V^{tot}$ & 10  \\   
		$P_{\rm th}/k$ (K cm$^{-3}$) & 5$\times$10$^7$-5$\times$10$^8$ \\
		$G_0^{\rm incident}$ (Mathis Unit) & 2$\times$10$^4$ \\
		UV field shape & Blackbody at 40,000K \\
		\hline 
		Transfer & Full line \\
		Cosmic rays (s$^{-1}$ per H$_2$) & 5$\times$10$^{-17}$ \\
		R$_V$ & 5.62 \\
		$N_{\rm H}$/E(B-V) (cm$^{-2}$) & 1.08$\times$10$^{22}$ \\ 
		Dust to gas ratio & 0.01 \\
		Grain size distribution & $\propto \alpha^{-3.5}$ \\
		min grain radius (cm) & 1$\times$10$^{-7}$ \\
		max grain radius (cm) & 3$\times$10$^{-5}$ \\
		\hline                                   
	\end{tabular}
\end{table}

To estimate the absolute intensity of the OH mid-IR lines, we consider the ratio of the OH lines to the H$_2$ 1-0 S(1) line at 2.12 $\mu$m predicted by the model multiplied by the H$_2$ 1-0 S(1) line intensity recently measured by the Keck Telescope \citep{Habart_2022} at an angular resolution (0.1") similar to that of the \textit{JWST}. Indeed, OH and H$_2$ lines originate roughly from the same region in the PDR (see Fig. \ref{fig:OH_H2_position}) so the intensity of these lines will be affected in the same way by geometry effects. Through the position of the \textit{JWST}/NIRSpec-IFU and MIRI-IFU mosaics of the ERS program, the 1-0 S(1) line was measured at an intensity of 8.7$\times$10$^{-4}$ erg cm$^{-2}$ s$^{-1}$ sr$^{-1}$ at the dissociation front. 

One need to correct the H$_2$ line intensity at 2.12 $\mu$m  for extinction due to the foreground dust and internal dust in the Bar itself.  As discussed 
in \cite{Habart_2022}, the H$_2$ line is expected to be about 56\% brighter in total (16\% for the foreground dust and 40\% for the internal dust). This is in agreement with the total extinction correction derived by \cite{Kaplan_2021}. This leads to an intensity corrected for the extinction of 1.4$\times$10$^{-3}$ erg cm$^{-2}$ s$^{-1}$ sr$^{-1}$. 
OH lines at longer wavelengths are mostly not affected by dust extinction.
It is only at shorter wavelengths that dust extinction will significantly attenuate the emission. The different OH/H$_2$ lines ratio and the estimated OH line intensity are presented in Table \ref{tab:OH_H2_ratio}.

\begin{table*}
	\centering
	\begin{tabular}{ccc}
		\hline\hline
		Models & $P_{\rm th}/k$ = 5$\times$10$^7$ K cm$^{-3}$ & $P_{\rm th}/k$ = 5$\times$10$^8$ K cm$^{-3}$ \\
		\hline
		OH(9.9 $\mu$m)/H$_2$(2.12 $\mu$m)  & 7.6$\times$10$^{-4}$ & 8.7$\times$10$^{-3}$ \\
		OH(10.8 $\mu$m)/H$_2$(2.12 $\mu$m)  & 8.1$\times$10$^{-4}$ & 9.2$\times$10$^{-3}$ \\
		\hline
		9.9 $\mu$m OH line intensity (erg cm$^{-2}$ s$^{-1}$ sr$^{-1}$) & 1.0$\times$10$^{-6}$ & 1.2$\times$10$^{-5}$ \\
		10.8 $\mu$m OH line intensity (erg cm$^{-2}$ s$^{-1}$ sr$^{-1}$) & 1.1$\times$10$^{-6}$ & 1.3$\times$10$^{-5}$ \\
		\hline\hline
		Integration time & Sensitivity at 9.9 $\mu$m (erg cm$^{-2}$ s$^{-1}$ sr$^{-1}$) - SNR = 5 & Sensitivity at 10.8 $\mu$m\\
		\hline
		111s & 1.4$\times$10$^{-6}$  & 1.2$\times$10$^{-6}$ \\
		15min & 5.0$\times$10$^{-7}$ & 4.3$\times$10$^{-7}$ \\
		\hline
	\end{tabular}
	\caption{OH/H$_2$(2.12 $\mu$m) line ratio and estimated OH line sum intensity of the quadruplet for different models. Sensitivity of the \textit{JWST} for a SNR of 5.}
	\label{tab:OH_H2_ratio}
\end{table*}

The line sensitivity of MIRI MRS (medium resolution spectroscopy) is about 1$\times$10$^{-6}$ erg cm$^{-2}$ s$^{-1}$ sr$^{-1}$ for a SNR of 5 and an integration time of 111 seconds, which is granted for the ERS program. The intensities for lines around 10 $\mu$m are about 1$\times$10$^{-6}$ erg cm$^{-2}$ s$^{-1}$ sr$^{-1}$ for the lower limit model at $P_{\rm th}/k$ =5$\times$10$^7$ K cm$^{-3}$ and around 1$\times$10$^{-5}$ erg cm$^{-2}$ s$^{-1}$ sr$^{-1}$ for the upper limit model at $P_{\rm th}/k$ =5$\times$10$^8$ K cm$^{-3}$ (see Table \ref{tab:OH_H2_ratio}). Both lines might be detected for the upper limit model only considering the estimated intensities. However, it is also possible to stack different lines to increase the SNR and detect them even in the lower limit model. Nevertheless the main limitation for detection of OH mid-IR lines in PDR is the small contrast between the strong continuum and the weak OH lines, as explained in the following section.

\subsubsection{Predicted spectrum with continuum and other lines}

\begin{figure}[!b]
	\centering
	\includegraphics[width=\linewidth]{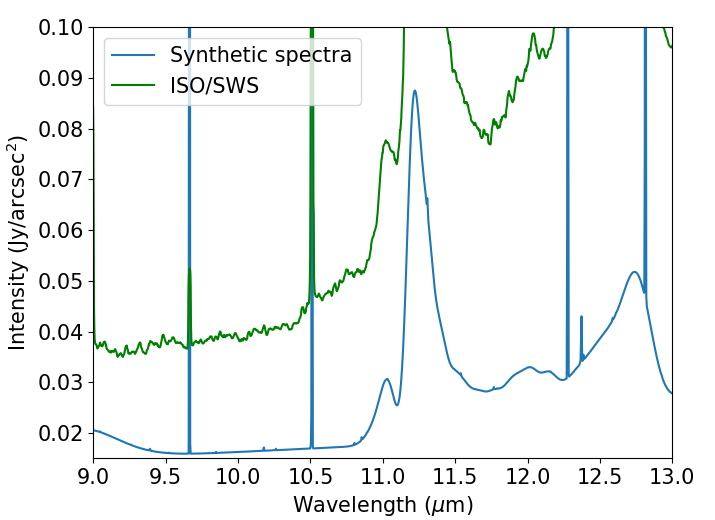}
	\caption{Synthetic spectrum at the H$^0$/H$_2$ dissociation front from the synthetic cube done for the future MIRI observation in the ERS program \citep{ERS_2022} (R$\sim$3000) and \textit{ISO}-SWS spectrum (R$\sim$1500).}
	\label{fig:region_spectra}
\end{figure}

To study the detectability of OH lines, we estimated the continuum and the other gas lines at the H$^0$/H$_2$ dissociation front where the OH lines emission peaks. OH lines could blend with other lines but most importantly could be lost in the continuum noise due to instrumental defaults such as fringing. Thus, it is necessary to have OH lines over the continuum greater than the noise expected on the continuum (predicted to be of order of a percent). To estimate the dust continuum and other lines, we have used the synthetic spectro-imaging cube from the ERS program. This cube was computed using five regions maps of the PDR and five template spectra of these regions\footnote{The Orion Bar synthetic spectra are available here \url{https://pdrs4all.org/seps/\#Orion-Bar-synthetic-spectra}}. The template spectra were determined using the PDR code for atomic and molecular lines contribution \citep{Le_Petit_2006}, \texttt{CLOUDY} for the ionized gas \citep{Ferland_1998}, the model \texttt{PAHTAT} for PAH emission \citep{Pilleri_2012} using the template spectra extracted by \cite{Foschino_2019} on \textit{ISO}-SWS data using machine learning algorithms, and finally the \texttt{THEMIS} dust model \citep{Jones2013,jones2017} with the radiative transfer code \texttt{SOC} \citep{Juvela2019} for the dust continuum emission following the approach of \cite{Schirmer_2022}. This dust model is based on \textit{Spitzer} and \textit{Herschel} observations in five photometric bands (3.6, 4.5, 5.8, 8, and 70 $\mu$m). Observations at 24 $\mu$m of the Orion Bar exist but are saturated.
We compare it to available observations that is \textit{ISO}/SWS observations centered on the PDR with a large beam (20") and \textit{Spitzer}-IRAC maps.  \textit{Spitzer}-IRS  spectrum are only available in the atomic region at the peak of the mid-IR dust continuum. The model reproduces well the \textit{Spitzer}-IRAC observations at 3.6 and 8 $\mu$m. However longward of 8 $\mu$m, the continuum in the synthetic spectrum is 3 times weaker than that measured in the \textit{ISO}-SWS spectrum (see Fig. \ref{fig:region_spectra}). 
Nevertheless, the \textit{ISO}-SWS spectrum does not spatially resolve the PDR and mixes the peak of the continuum observed in the atomic zone with \textit{Spitzer}/IRAC and the peak of the lines of H$_2$ and OH expected at the dissociation front. \textit{JWST} will spatially resolve the different PDR layers, allowing us to properly verify the model and determine if the continuum is actually underestimated for $\lambda$ > 8 $\mu$m.

Fig. \ref{fig:region_spectra} displays the spectrum at the H$^0$/H$_2$ dissociation front derived from the synthetic cube. We focus here on the wavelength range 9 to 13 $\mu$m because as seen in Fig. \ref{fig:spectre_OH} this is the domain where OH lines are the brightest. Moreover, longward of $15~\mu$m, OH lines might also be excited by other mechanisms such as chemical pumping by O+H$_2$ with ro-vibrationally excited H$_2$ \citep[A. Zanchet, private communications and primarily results from][]{Maiti_2003,Braunstein_2004,Weck_2006} or with the first excited state of oxygen O($^1$D) \citep{Liu_2000} and not only from water photodissociation. 

\subsubsection{Line-continuum ratio}
\label{line-continuum}
\begin{figure}
	\centering
	\includegraphics[width=\linewidth]{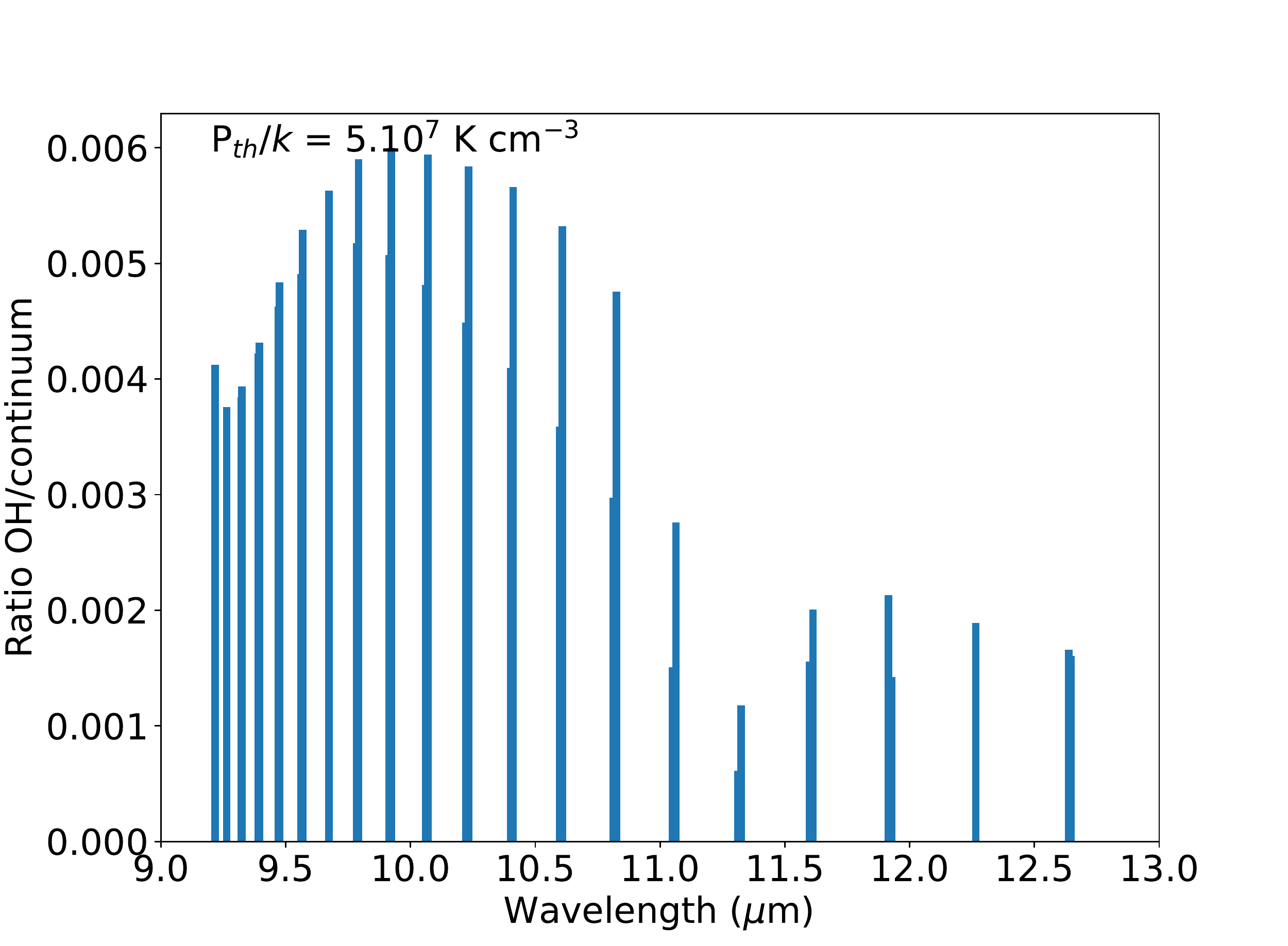}
	\includegraphics[width=\linewidth]{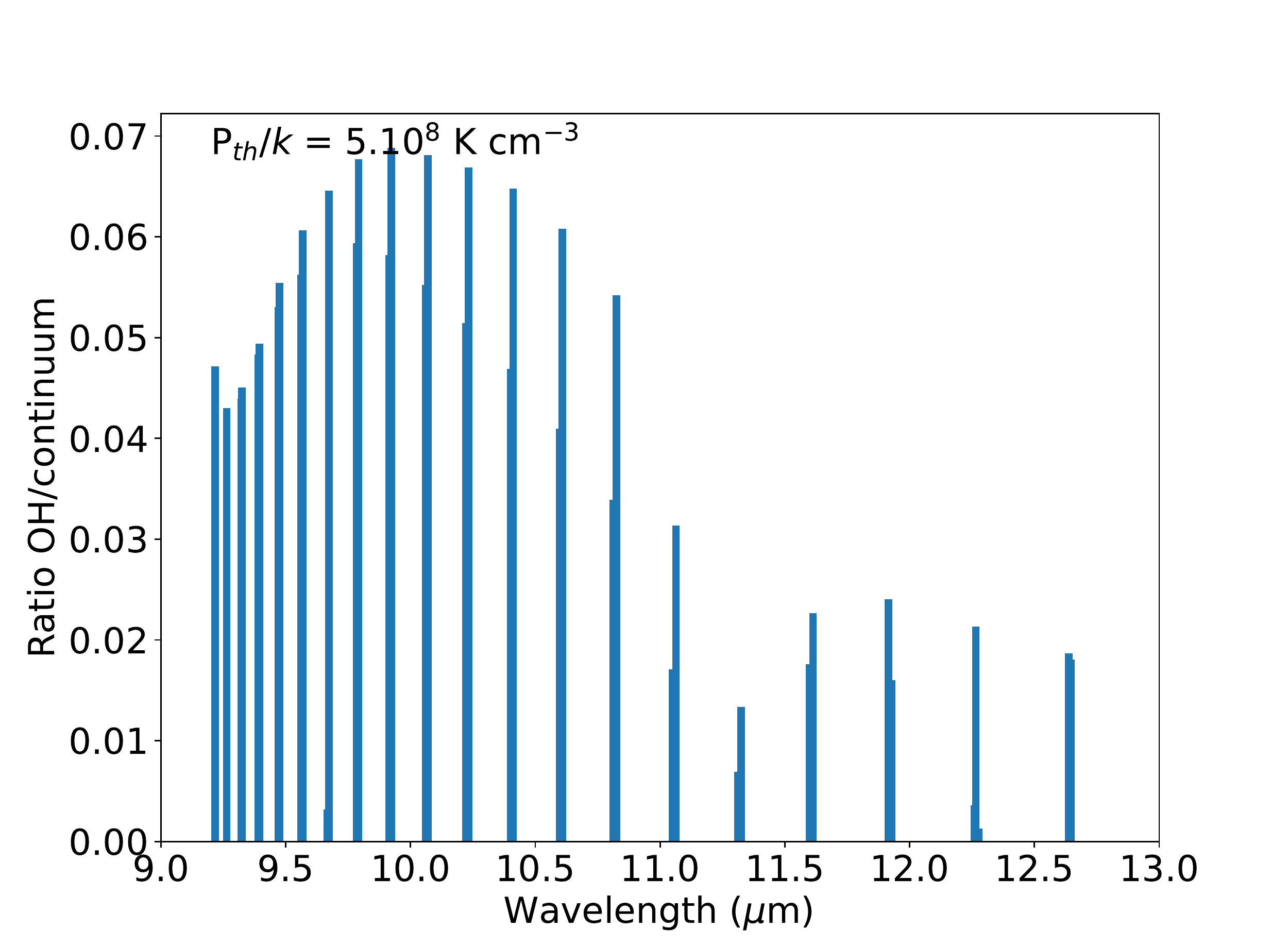}
	\caption{Ratio of OH lines over the continuum. (Top panel) $P_{\rm th}/k$ = 5.10$^7$ K cm$^{-3}$ and $G_0^{\rm incident}$ = 10$^4$. (Bottom panel) $P_{\rm th}/k$ = 5.10$^8$ K cm$^{-3}$ and $G_0^{\rm incident}$ = 10$^4$. The main difference from the variation of the OH line intensities seen in Fig. \ref{fig:spectre_OH} is at 11.3 $\mu$m where the ratio drops drastically due to the aromatic band (at 11.3 $\mu$m).}
	\label{fig:ratio_OH_continuum}
\end{figure}

Figure \ref{fig:ratio_OH_continuum} shows OH line over continuum ratio for models with different thermal pressure. In order to compute the line to continuum ratio, OH line spectra were calculated by considering the integrated intensities divided by the frequency width equal to the spectral resolution element. Then, the spectra were ratioed to the simulated spectrum including the continuum and other gas lines as described in the previous sub-section. The ratio is much higher for the $P_{\rm th}/k$ = 5.10$^8$ K cm$^{-3}$ model than for the $P_{\rm th}/k$ = 5.10$^7$ K cm$^{-3}$ model as expected from the analysis of Sect. \ref{pressure}. The line-to-continuum variation results from the OH line intensities distribution as seen in Fig. \ref{fig:spectre_OH} and the shape of the dust emission continuum, aromatic bands, and bright lines in this region as seen in Fig. \ref{fig:region_spectra}. The overall shape is similar for both models because as explained in Sect. \ref{OH_midIR_prediction}, the relative intensities of the intra-ladder lines depend only on the spectral shape of the UV field which here is the same for both models. The ratio of OH lines over the continuum is maximum around 10 $\mu$m. This is a result of a combined effect between OH line intensities reaching its maximum around 10.8 $\mu$m and continuum increasing progressively after 10 $\mu$m. The lines around 10 $\mu$m are thus those with the highest probability of detection with the \textit{JWST}. The value of the maximum ratio varies from 0.6 \% for a lower limit model at $P_{\rm th}/k$ = 5$\times$10$^7$ K cm$^{-3}$ to 7\% for an upper limit model at $P_{\rm th}/k$ = 5$\times$10$^8$ K cm$^{-3}$. The continuum could be underestimated after 8 $\mu$m which can lead to smaller ratios.

Thus the detection should be possible for the upper limit model at $P_{\rm th}/k$ = 5$\times$10$^8$ K cm$^{-3}$, while it will be more challenging for the lower limit model at $P_{\rm th}/k$ = 5$\times$10$^7$ K cm$^{-3}$.

\subsubsection{Possible blending with other lines}

\begin{figure}
	\centering
	\includegraphics[width=\linewidth]{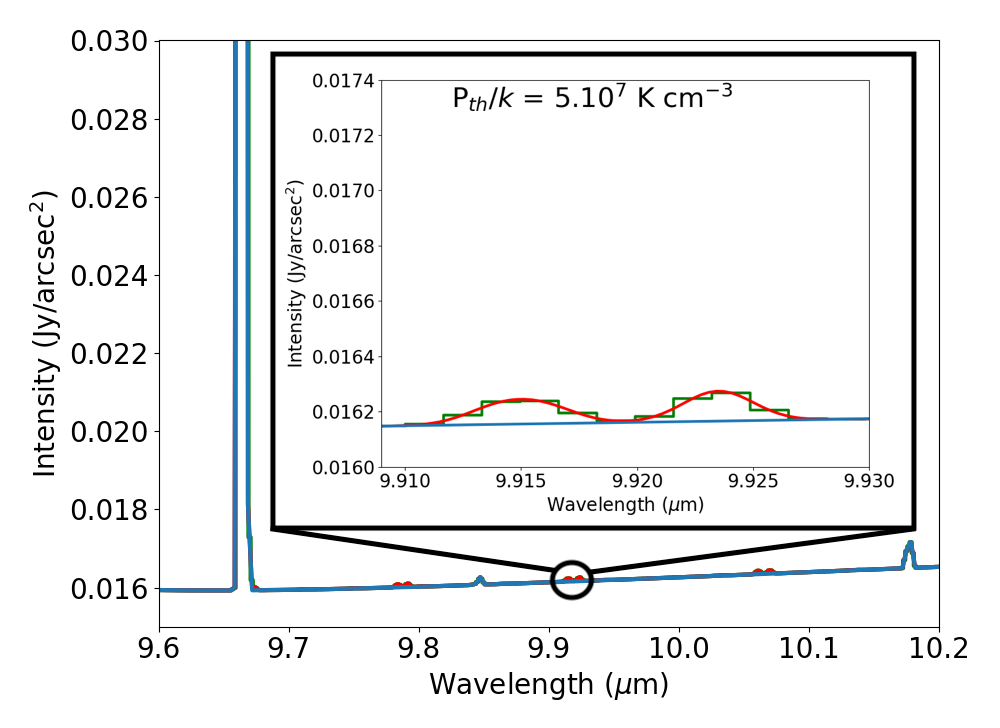} \includegraphics[width=\linewidth]{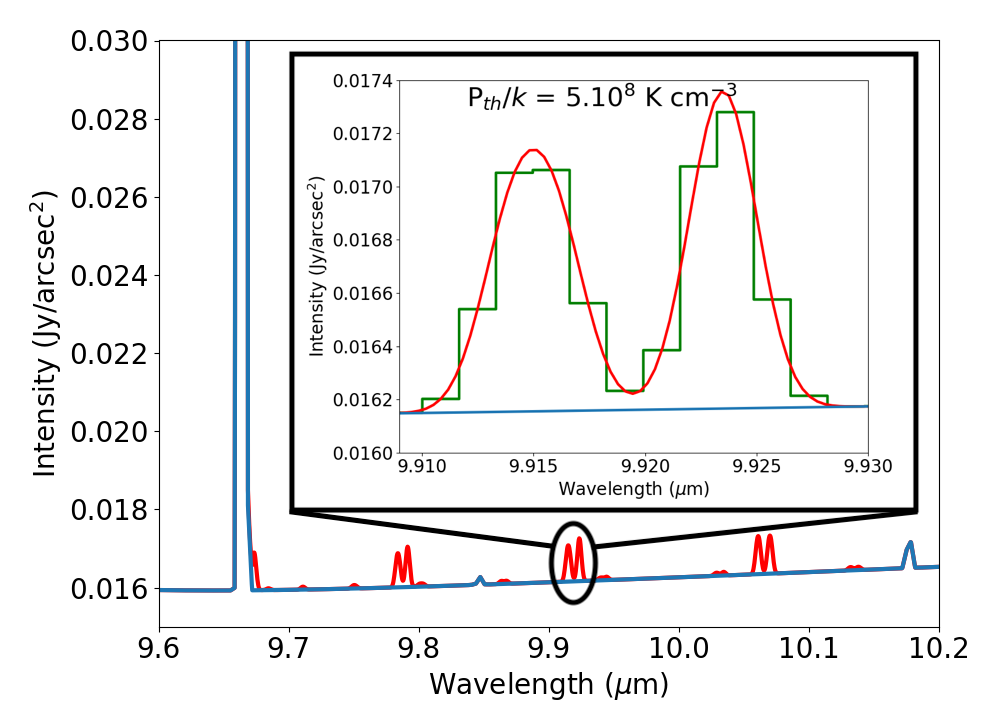}
	\caption{Spectrum near the dissociation front at the position chosen for MIRI IFU observation with the ERS \citep{ERS_2022} around 10 $\mu$m (blue) with the addition of OH lines (red) for a spectral resolving power R = 3000. The line at 9.67 $\mu$m is the H$_2$ 0-0 S(3). The sampling, represented with the green line, is 1/2 times the element of resolution around 10 $\mu$m \citep[as Fig. 7 from][]{Glasse_2015}. It is about 1.7$\times$10$^{-3}$ $\mu$m near 10 $\mu$m for R =3000.  (Top panel) $P_{\rm th}/k$ = 5.10$^7$ K cm$^{-3}$ and $G_0^{\rm incident}$ = 10$^4$. (Bottom panel) $P_{\rm th}/k$ = 5.10$^8$ K cm$^{-3}$ and $G_0^{\rm incident}$ = 10$^4$. The 9.9 $\mu$m quadruplet is only partially resolved with a resolving power of 3000. Only two peaks are visible instead of four expected (at 9.914 $\mu$m, 9.916 $\mu$m, 9.923 $\mu$m and 9.924 $\mu$m).}
	\label{fig:region_OH}
\end{figure}

There are also intense lines such as H$_2$ rotational lines (0-0 S(2) and 0-0 S(3)), aromatic bands or lines from the ionized gas (NeII at 12.8 $\mu$m, SIV at 10.5 $\mu$m, ...) which could blend with OH lines. Fig. \ref{fig:region_OH} displays a zoom on the 10 $\mu$m zone of the spectrum where OH lines are added with a resolving power of 3000. This specific wavelength domain was chosen because this is where the line over continuum ratio is the highest (see Fig. \ref{fig:ratio_OH_continuum}). On this figure, the line seen at 9.67 $\mu$m is the H$_2$ 0-0 S(3) line. This figure highlights the fact that OH lines are drastically less intense than other lines. Moreover, the 9.9 $\mu$m quadruplet is only partially resolved with a resolving power of 3000 (close to the \textit{JWST} resolving power at this wavelength). Indeed, only two peaks are visible instead of four expected (at 9.914 $\mu$m, 9.916 $\mu$m, 9.923 $\mu$m and 9.924 $\mu$m). However, this figure shows that in the region where the line over continuum ratio is the highest, no other lines should blend with OH lines. \\

In summary, this study shows that in terms of intensity, it is probable OH lines could be detectable as long as the thermal pressure is not too low. The main difficulty to detect those lines lies in the line-continuum ratio. Indeed, as seen in Sect. \ref{line-continuum}, even in the best case scenario, the line to continuum ratio will not exceed 1-7\%. The possible detectability of OH lines relies on the low level of noise. The noise has to be below the ratio line-continuum to enable detection.

\subsection{Application to other environments}
\subsubsection{Other interstellar PDRs}
Figure \ref{fig:I30_Pgas_total} and Sect. \ref{results_grid} show that OH mid-IR lines can only be detected in very illuminated PDRs with high pressure due to the presence of high temperature. In particular, OH mid-IR lines are expected to be too weak in regions such as the Horsehead nebula because the thermal pressure, hence gas temperature, is too low. Regions such as NGC 7023 might be a candidate to produce bright enough OH lines but these would be at the limit of detection (about 1$\times$10$^{-6}$ erg cm$^{-2}$ s$^{-1}$ sr$^{-1}$ for an integration time of 144 seconds a SNR of 5 which is granted for the GTO 1192). What is highlighted in this figure is that very high pressure $P_{\rm th}/k \gtrsim$ 5.10$^7$ K cm$^{-3}$ and UV field intensity $G_{0} >$ 10$^3$ are necessary to produce supposedly detectable OH mid-IR lines. Due to this result, the Orion Bar is the best candidate to observe them.

\subsubsection{Proplyds}

PDRs at the edge of dense molecular clouds are not the only objects where OH lines might be detected by \textit{JWST}. Our thermochemical modeling shows that, as a rule of thumb, denser irradiated environments (high thermal pressure) result in stronger mid-IR OH line fluxes. This result is in line with the previous detections of bright OH lines with \textit{Spitzer}-IRS in several proto-planetary disks \citep[][ Tabone et al., in prep]{Carr_2014} and strong protostellar shocks \citep{Tappe_2008, Tappe_2012}.
\textit{JWST} programs dedicated to interstellar PDRs will encompass proplyds in their field of view \citep[e.g. ERS PDRs4All,][]{ERS_2022}. Because they correspond to very dense clumps of gas \citep[$n_H > 10^{8}$cm$^{-3}$][]{Champion_2017}, and based on the previous detection of non-externally irradiated disks, OH mid-IR lines are expected to be well detected with \textit{JWST}. In these objects, OH mid-IR emission will still be directly related to the amount of H$_2$O photodissociated per unit time. However, detailed interpretation of the line flux requires dedicated modeling of proplyds, which is beyond the scope of the present paper.

\section{Conclusion}
\label{conclusion}
In this work, we explored the potential of OH mid-IR lines for the study of interstellar PDRs. In order to achieve this goal, we amended the Meudon PDR Code to include prompt emission induced by H$_2$O photodissociation in the 114-143 nm UV band and new state-specific formation rate of OH, and analyse a grid of model.

The main conclusions of this study are :
\begin{enumerate}
	\item OH mid-IR emission is confined to a thin exposed layer close to the H$^0$/H$_2$ transition where H$_2$O is formed by neutral-neutral reactions and actively photodissociated.
	\item OH mid-IR lines are directly proportional to the column density of water photodissociated in the $114-143~$nm range. Since water requires high temperature to form, OH mid-IR lines are very sensitive  to the temperature at the H$^0$/H$_2$ transition. In particular, we predict that OH mid-IR lines are brighter in regions with high thermal pressure.
	\item OH mid-IR lines are less dependent on the strength of the incident UV field for $G_0^{incident}$ > 10$^3$. When the incident UV field increases, the H$^0$/H$_2$ transition shifts deeper into the cloud but the temperature and local UV field (and thus water abundance) stay rather constant, explaining the rather small impact on OH lines.
	\item OH lines are predicted to be detectable with \textit{JWST} only in highly illuminated PDRs ($G^{incident}_0$ > 10$^3$) with high pressure ($P_{\rm th}/k$ > 5$\times$10$^7$ K cm$^{-3}$). Detection might then be possible in the Orion Bar but not in the Horsehead Nebula. The low line-to-continuum ratio might also be a major limitation for the detection due to instrumental effect such as fringes. 
	
\end{enumerate}

To conclude, our work demonstrates that OH mid-IR lines are a promising tool to study the physical processes in PDRs. In particular, OH mid-IR lines constitute an indirect but sensitive diagnostic of the temperature at the H$^0$/H$_2$ transition, a parameter that is highly uncertain in PDR models \citep{Rollig2007}. Spatially resolved observations of mid-IR OH and H$_2$ rotational lines will therefore be key to test PDR models and better calibrate the correlation between OH emission and temperature, and study in details oxygen chemistry in irradiated environments. For unspatially resolved observations of PDRs, where H$_2$ emission gives only an average estimate of the temperature, OH would then be a unique diagnostic to access the temperature around the H$^0$/H$_2$ transition. We also note that in this study, we focused only on the prompt emission of OH induced by H$_2$O photodissociation at shortwavelength. Prompt emission induced by water photodissociation longward of $143~$nm, which excites ro-vibrational lines in the near-IR, as well as chemical pumping by O+H$_2$, which excites mid-IR lines longward of $\simeq 15~\mu$m, if properly modeled, can also bring strong complementary constraints on the physical and chemical processes in dense PDRs.

\begin{acknowledgements}
	The authors wish to thank John H. Black for insightful comments on the manuscript and Alexandre Zanchet for fruitful discussions about chemical pumping of OH.
	This work was partially supported by the Programme National “Physique et Chimie du Milieu Interstellaire” (PCMI) of CNRS/INSU with INC/INP and co-funded by CNES. B.T. is a Laureate of the Paris Region fellowship program, which is supported by the Ile-de-France Region and has received funding under the Horizon 2020 innovation framework program and Marie Sklodowska-Curie grant agreement No. 945298.
\end{acknowledgements}

\bibliographystyle{aa}
\bibliography{biblio}

\begin{appendix}
	\section{H$_2$O Chemistry}
	
	\label{appendix_chemistry}
	\subsection{Neutral-neutral reactions}
	
	\label{neutral_reaction}

	\begin{table*}[]
		\centering
		\begin{tabular}{ccc}
			\hline
			Reaction & Reaction rate & Reaction rate coefficient \\
			\hline\hline
			O + H$_2$ = OH + H & $v_1 = k_1 n(O)n(H_2)$ & $k_1 = 2.22\times10^{-14}\times\left(\frac{T}{300}\right)^{3.75}\times \exp(-2401/T)$ \\\hline
			OH + H$_2$ = H$_2$O + H & $v_2 = k_2 n(OH)n(H_2)$   & $k_2 = 2.22\times10^{-12}\times\left(\frac{T}{300}\right)^{1.43}\times \exp(-1751/T)$\\\hline
			OH + photon = O + H & $v_{\phi_1} = k_{\phi_1} G_0n(OH)$ & $k_{\phi_1} = 1.32\times10^{-10}$ \\\hline
			H$_2$O + photon = OH + H & $v_{\phi_2} = k_{\phi_2}G_0 n(H_2O)$ & $k_{\phi_1} = 2.09\times10^{-10}$\\\hline
		\end{tabular}
		\caption{Thermal rate coefficient for chemical reactions for O+H$_2$ \citep{Veselinova_2021} and OH+H$_2$ \citep{Agundez_2010}, and the photodissociation rate of OH and H$_2$O \cite{Heays_2017}.}
		\label{tab:kinetic}
	\end{table*}
	
	In the warm region, at low $A_V$ ($A_V$ $\le$ 1), H$_2$O is formed via neutral-neutral reactions:
	
	\begin{center}
		\ce{O <=>[H_2][UV] OH <=>[H_2][UV] H_2O}
	\end{center}
	Assuming that the abundance of total oxygen (x$_0 \simeq 3.19 \times 10^{-4}$) is the sum of the abundance of atomic oxygen, water, and hydroxyl, we get a steady-state abundance of H$_2$O of:
	
	\begin{equation}
		x(H_2O) = \frac{x_O}{1+\frac
			{k_{\phi_1}k_{\phi_2}}{k_1k_2 x(H_2)^2} \left(\frac{G_0}{n_H}\right)^{2} + \frac
			{k_{\phi_2}}{k_2x(H_2)} \left(\frac{G_0}{n_H}\right)} \\
	\end{equation}
	The rate coefficients used in this formula are presented Table \ref{tab:kinetic}. In the appendix, we use the thermal rates  from \cite{Agundez_2010} and \cite{Veselinova_2021}. This formula is a generalization of \citet{Kristensen2017} \citep[see also][]{vanDishoeck2021} and shows that H$_2$O abundance in warm and dense environments depends primarily on G$_0$/n$_H$, $T_{\rm K}$, and H$_2$ abundance.
	
\begin{figure}[!h]
		\centering
		\includegraphics[width=\linewidth]{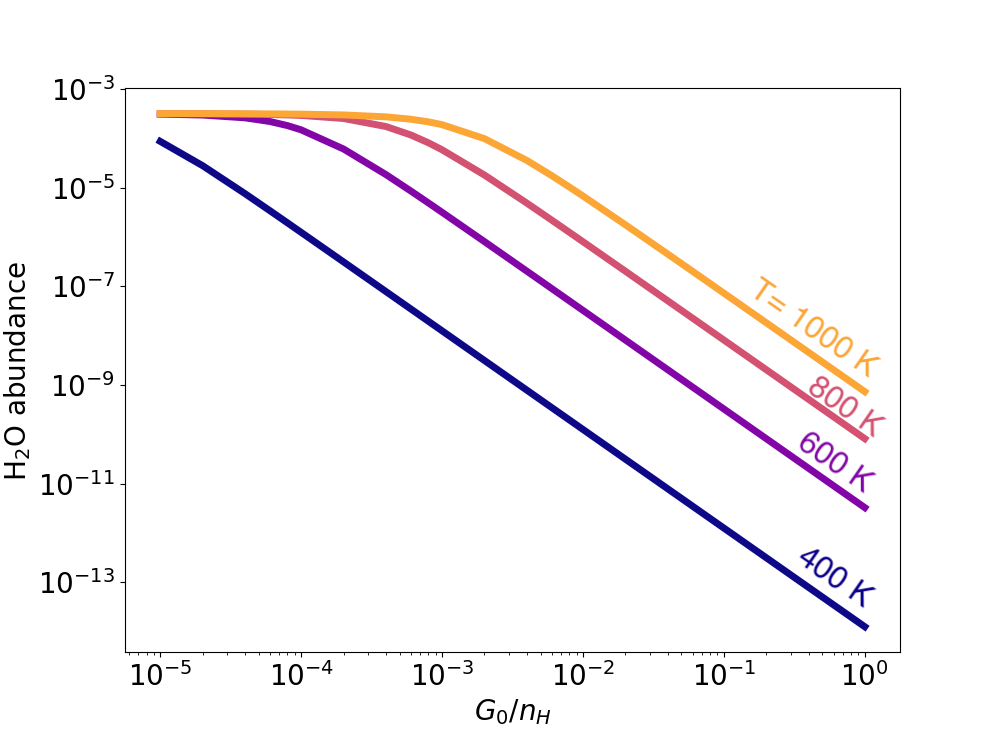}
		\caption{H$_2$O abundance from the neutral-neutral reactions as a function of $G_0^{\rm incident}$/$n_{\rm H}$ for different value of temperature.}
		\label{fig:abondance_annexe}
	\end{figure}

	The abundance is displayed Fig. \ref{fig:abondance_annexe} as a function of $G_0$/$n_{\rm H}$ for different temperature. The abundance increases dramatically with temperature in the $T_{\rm K}= 400-1000~$K range whereas it is inversely proportional to $G_0$/$n_{\rm H}$. This shows that a small change in the temperature results in a large variation in H$_2$O abundance and therefore in OH mid-IR emission whereas a change in incident UV radiation field has a somewhat smaller impact. 
	
	We also show in Fig. \ref{fig:abondance_annexe_3} that our analytical model reproduces well the abundance profile computed by the Meudon PDR code, confirming that, in the case of H$_2$ + O reactions, thermal reaction rates can be use to analyse, at least qualitatively, the role of the physical conditions on the amount of H$_2$O and therefore on OH mid-IR emission.
	
	We note that in this appendix, we only use the thermal coefficient rate assuming H$_2$ levels follow a Boltzmann distribution. H$_2$ levels distribution can be different in PDRs, in particular due to UV pumping. However, using the state-specific chemistry from \cite{Veselinova_2021} only increase the abundance of H$_2$O by a factor 2.
	
	\subsection{Ion-neutral reactions}
	
	\label{H2O_abundance_2}
	
	In colder regions, at higher $A_V$ ($A_V$ $\ge$ 1), H$_2$O is formed via ion-neutral reactions. In atomic regions, the ion-neutral route is:
	
	\ce{H ->[\zeta] H^+ ->[O] O^+ ->[H_2] OH^+ ->[H_2] H_2O^+ ->[H_2] H_3O^+ ->[e^-/\eta] H_2O ->[UV] OH}\\
	$\zeta$ is the cosmic rays ionisation rate and $\eta$ is the branching ratio of the electronic recombination forming H$_2$O ($\eta$ $\sim$ 20 \%). Indeed, the electronic recombination of H$_3$O$^+$ can also lead to OH. We further assume that this series of reactions leading to H$_3$O$^+$ has an efficiency $\epsilon$ ($\epsilon$ $\sim$ 15\%):\begin{equation}
		x(H_2O) = \epsilon\eta\frac{\zeta}{k_{\phi_2}G_0}x(H)
	\end{equation}
	
	In molecular regions, the ion-neutral route is slightly different:
	
	\ce{H_2 ->[\zeta] H_2^+ ->[H_2] H_3^+ ->[O] OH^+ ->[H_2] H_2O^+ ->[H_2] H_3O^+ ->[e^-/\eta] H_2O ->[UV] OH}\\
	
	Thus in molecular regions, we have:\begin{equation}
		x(H_2O) = \epsilon\eta\frac{\zeta}{k_{\phi_2}G_0}x(H_2).
	\end{equation}
	In fig. \ref{fig:abondance_annexe_3}, we show that is analytical formula reproduces well the abundance profile of H$_2$O in the cold and shielded region of the PDR where the neutral-neutral formation route is inefficient. However, as mentionned in Sec. \ref{Meudon_PDR}, only gas-phase chemistry is taken into account into the Meudon PDR Code for water formation. In those regions, formation of solid-H$_2$O and subsequent photo-desorption back to the gas phase may alter the position and amplitude of the peak \citep{Hollenbach_2009,Putaud_2019}.
	
	\label{summary_appendix}
	\begin{figure}[!h]
		\centering
		\includegraphics[width=\linewidth]{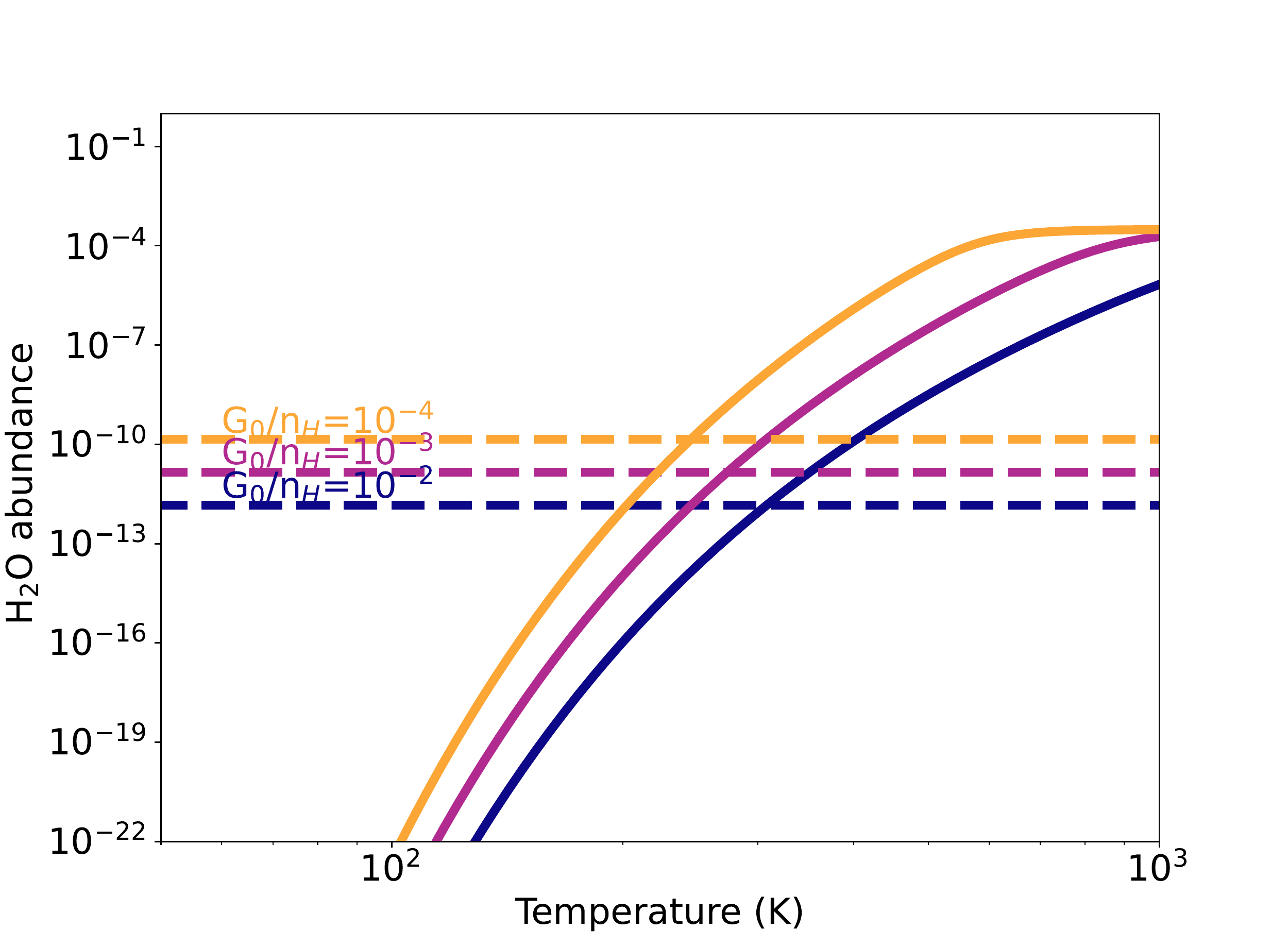}
		\caption{H$_2$O abundance as a function of temperature for different value of $G_0$/$n_{\rm H}$. The dashed lines are the abundances calculated for the ion-neutral route and the solid lines are the abundances calculated for the neutral-neutral route. This figure highlights the transition temperature between ion-neutral and neutral-neutral route.}
		\label{fig:abondance_annexe_2}
	\end{figure}
	
	\begin{figure}[!h]
		\centering
		\includegraphics[width=\linewidth]{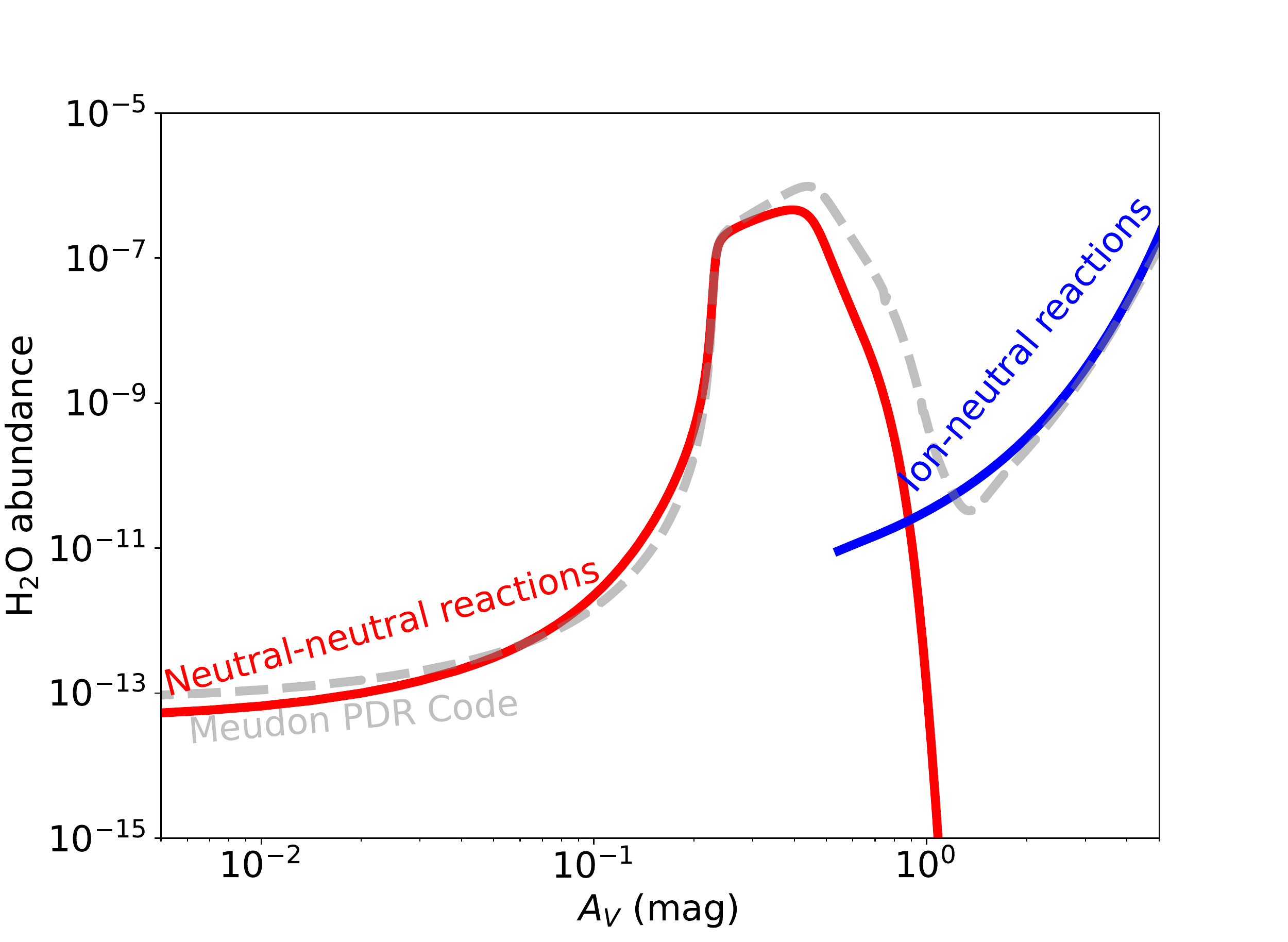}
		\caption{Comparison between the Meudon PDR Code for a model at $P_{\rm th}/k$ = 10$^8$ K cm$^{-3}$ and G$^{incident}_0$ = 10$^4$ and the analytic calculation of H$_2$O abundance. The difference between the model and the analytic calculation is due to the fact we do not take into account other formation processes.
		}
		\label{fig:abondance_annexe_3}
	\end{figure}
	
	\section{OH chemistry}
	
	\label{appendix_chemistry_OH}
	
	\begin{figure}
		\centering
		\includegraphics[width=\linewidth]{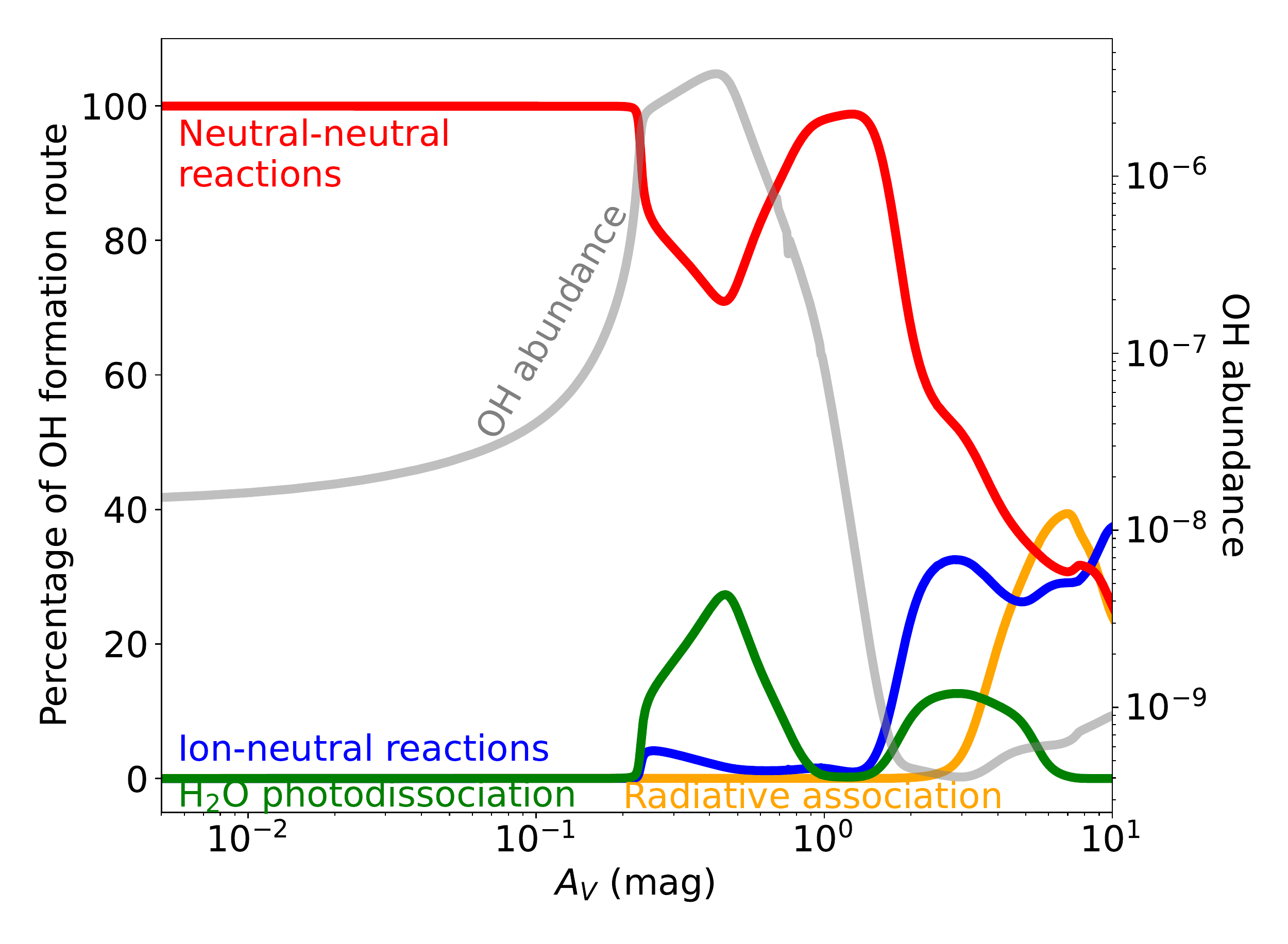}
		\caption{Percentage of the different formation routes of OH as a function of the visual extinction for the model $P_{\rm th}/k$ = 10$^8$ K cm$^{-3}$ and $G_0^{\rm incident}$ = 10$^4$. The blue line represents the percentage of the ion-neutral formation route, and the red line represents the percentage of the neutral-neutral formation route that requires high temperatures ($T_{\rm K} \gtrsim 300~$K). The green line represents the percentage of the formation of OH via H$_2$O photodissociation. The orange line represents the percentage of the formation of OH via radiative association.}
		\label{fig:OH_chemistry}
	\end{figure}
	
	Fig. \ref{fig:OH_chemistry} displays the different formation routes of OH. It shows that in most of the cloud, the neutral-neutral reactions dominate. There is a significant contribution of H$_2$O photodissociation at the OH abundance peak.

\end{appendix}

\end{document}